\documentclass[10pt,journal,compsoc]{IEEEtran}

\ifCLASSOPTIONcompsoc
  \usepackage[nocompress]{cite}
\else
  \usepackage{cite}
\fi

\ifCLASSINFOpdf
\else
\fi

\usepackage{algorithmic}
\usepackage{graphicx}
\usepackage{textcomp}
\usepackage{xcolor}
\usepackage{hyperref}
\usepackage{url}
\usepackage{multirow}
\usepackage{balance}
\usepackage{url}
\usepackage{stfloats}
\usepackage{booktabs}
\usepackage{makecell}
\usepackage{enumitem}
\usepackage{array}
\usepackage{tabularx}
\usepackage{booktabs}
\usepackage{xspace}
\usepackage{xfrac}
\usepackage{tabu}
\usepackage{array}
\usepackage{diagbox}
\usepackage[normalem]{ulem}
\usepackage{threeparttable}

\usepackage{algorithm}
\usepackage{algorithmic}

\AtBeginDocument{%
  \providecommand\BibTeX{{%
    \normalfont B\kern-0.5em{\scshape i\kern-0.25em b}\kern-0.8em\TeX}}}

\usepackage{xcolor}
\usepackage{enumitem}
\usepackage{amsmath,amssymb}

\newcommand{\revised}[1]{\textcolor{black}{#1}}

\usepackage{graphicx}
\usepackage{hyperref}
\usepackage{url}

\usepackage{comment}
\usepackage{colortbl}
\usepackage{graphicx, subfigure}
\usepackage[most]{tcolorbox}

\usepackage[utf8]{inputenc}
\usepackage{siunitx}
\usepackage{hyperref}
\usepackage{url}
\usepackage{listings}
\usepackage{xcolor}
\usepackage{soulutf8}
\usepackage{tcolorbox}
\usepackage{tikz}

\usepackage{svg}
\usepackage{amsmath,amssymb}
\usepackage{enumitem}
\usepackage{float}
\usepackage{adjustbox}
\usepackage{soul}
\usepackage{xcolor}
\usepackage{ifthen}

\usepackage[justification=centering]{caption}

\usepackage{pifont}

\newboolean{showcomments}
\setboolean{showcomments}{true}
\ifthenelse{\boolean{showcomments}}
 {\newcommand{\mynote}[2]{
      \fbox{\bfseries\sffamily\scriptsize#1}
        {\small$\blacktriangleright$\textsf{\emph{#2}}$\blacktriangleleft$}}}
        { \newcommand{\mynote}[2]{}}

\usepackage{xspace}
\newcommand{\toolname}{{LLM4PatchCorrect}\xspace}

\definecolor{LightCyan}{rgb}{0.88,1,1}
\definecolor{Gray}{gray}{0.9}
\newcolumntype{g}{>{\columncolor{LightCyan}}r}
\newcolumntype{b}{>{\columncolor{Gray}}c}

\begin{document}

\title{Automatic Patch Correctness Assessment with Large Language Model}

\author{Xin~Zhou,~\IEEEmembership{}
        Bowen~Xu,~\IEEEmembership{}
        Kisub~Kim,~\IEEEmembership{}
        DongGyun~Han,~\IEEEmembership{}
        Hung~Huu~Nguyen,~\IEEEmembership{}
        Thanh~Le-Cong,~\IEEEmembership{}
        Junda~He,~\IEEEmembership{}
        Bach~Le,~\IEEEmembership{}      and~David~Lo,~\IEEEmembership{Fellow, IEEE}
\IEEEcompsocitemizethanks{
\IEEEcompsocthanksitem X. Zhou, K. Kim, H. H. Nguyen, J. He and D. Lo are with the School of Computing and Information Systems, Singapore Management University, Singapore. \protect \\
E-mail: \{xinzhou.2020, huuhungn, jundahe, davidlo\}@smu.edu.sg and falconlk00@gmail.com.
\IEEEcompsocthanksitem B. Xu is with the Department of Computer Science College of Engineering, North Carolina State University, United States (USA).\protect\\
E-mail: bxu22@ncsu.edu
\IEEEcompsocthanksitem D. Han is with the Department of Computer Science, Royal Holloway, University of London, United Kingdom (UK).\protect\\
E-mail: donggyun.han@rhul.ac.uk.
\IEEEcompsocthanksitem T. Le-Cong and B. Le is with the School of Computing and Information Systems, The University of Melbourne, Melbourne, Australia.\protect\\
E-mail: congthanh.le@student.unimelb.edu.au and\\ bach.le@unimelb.edu.au.}
}

\markboth{Journal of \LaTeX\ Class Files,~Vol.~xx, No.~x, xx~xxxx}%
{Shell \MakeLowercase{\textit{et al.}}: Bare Demo of IEEEtran.cls for Computer Society Journals}

\IEEEtitleabstractindextext{%
\begin{abstract}
Automated Program Repair (APR) techniques have shown more and more promising results in fixing real-world bugs. 
Despite the effectiveness, APR techniques still face an overfitting problem: a generated patch can be incorrect although it passes all tests.
It is time-consuming to manually evaluate the correctness of generated patches that can pass all tests.
To address this problem, many approaches have been proposed to automatically assess the correctness of patches generated by APR techniques. These approaches are mainly evaluated within the cross-validation setting.
However, for patches generated by a new or unseen APR tool, users are implicitly required to manually label a significant portion of these patches (e.g., 90\% in 10-fold cross-validation) in the cross-validation setting before inferring the remaining patches (e.g., 10\% in 10-fold cross-validation).
To mitigate the issue, in this study, we propose \textbf{\toolname}, the patch correctness assessment by adopting a large language model for code. Specifically, for patches generated by a new or unseen APR tool, \toolname does not need labeled patches of this new or unseen APR tool for training but directly queries the large language model for code to get predictions on the correctness labels without training.
In this way, \toolname can reduce the manual labeling effort when building a model to automatically assess the correctness of generated patches of new APR tools.
To provide knowledge regarding the automatic patch correctness assessment (APCA) task to the large language model for code, \toolname leverages bug descriptions, execution traces, failing test cases, test coverage, and labeled patches generated by existing APR tools, before deciding the correctness of the unlabeled patches of a new or unseen APR tool. Additionally, \toolname prioritizes labeled patches from existing APR tools that exhibit semantic similarity to those generated by new APR tools, enhancing the accuracy achieved by \toolname for patches from new APR tools.
Our experimental results showed that \toolname can achieve an accuracy of 84.4\% and an F1-score of 86.5\% on average although no labeled patch of the new or unseen APR tool is available.
In addition, our proposed technique outperformed the prior state-of-the-art by a large margin.

\end{abstract}

\begin{IEEEkeywords}
Automatic patch correctness assessment, Large language models of code, In-context learning
\end{IEEEkeywords}}

\maketitle

\IEEEdisplaynontitleabstractindextext

\IEEEpeerreviewmaketitle

\section{Introduction}
\label{sec:introduction}

Automated Program Repair (APR) has gained increasing attention and diverse APR tools have been proposed~\cite{Goues2012GenProgAG, Long2015StagedPR, Le2016HistoryDP,  Xin2017LeveragingSC, Jiang2018ShapingPR, Liu2018LSRepairLS,Liu2018MiningSF, Lin2020UnderstandingTN, Qin2021OnTI, Martin2020AutomaticSR, Liu2019TBarRT,Zhu2021ASE, Jiang2021CURECN}.
Despite the significant improvements achieved in APR, existing APR tools still face a long-standing challenge: the overfitting problem~\cite{Long2016AnAO, Le2017OverfittingIS, Le2019OnRO, Wang2019HowDI, Nilizadeh2021ExploringTT}.
Due to the absence of strong program specifications, APR tools often use test cases to validate whether a generated patch is correct or not. 
However, passing all the existing test cases does not ensure that the patch is indeed correct and there is no guarantee that the patch can actually repair the program. 
A generated patch is considered ``overfitting'' if it passes all the available test cases while it is still incorrect with respect to the intended program specification.

Identifying overfitting patches is crucial for the APR tool adoption in practice.
Suppose Bob is a practitioner who is keen to use advanced APR tools.
There exist multiple approaches he can employ, and each produces many patches. 
However, recent studies~\cite{Le2018OverfittingIS,Qi2015AnAO} demonstrate that APR tools could generate more overfitting patches than correct ones, showing a high false positive rate.
In addition, researchers have revealed that high false positive rates may deliver dissatisfaction and distrust to developers on automatic SE tools such as static analysis~\cite{Johnson2013WhyDS} and fault localization~\cite{Kochhar2016PractitionersEO}.
This indicates that APR tools can disappoint Bob by wasting his time with wrong (i.e., overfitting) patches. 
Thus, it is important to detect and reduce the false positives (i.e., overfitting patches), especially for the generate-and-validate APR approaches in practice~\cite{tian2020evaluating}.

To address this issue, many approaches~\cite{Tan2016AntipatternsIS, xiong2018identifying, Yang2017BetterTC, tian2020evaluating,  Ye2022AutomatedCO, Ye2021AutomatedPA, Wang2020AutomatedPC, Xin2017IdentifyingTP, Wen2018ContextAwarePG, Xin2017LeveragingSC, Ye2022AutomatedCO, Le2017S3SA, cache, quatrain, le2023invalidator} have been proposed to conduct automatic patch correctness assessment (APCA).
Lin et al.~\cite{cache} categorized the existing APCA approaches into two categories: 
(1) dynamic approaches which are based on running/executing the tests and (2) static approaches which are built on top of source code patterns or features.
In general, dynamic approaches perform correctness assessment by either augmenting test cases using automated test generation tools such as Randoop~\cite{Pacheco2007RandoopFR} or collecting the runtime information for analysis.
On the other hand, static approaches extract code patterns or features to decide the correctness.  
Despite the promising results, both of them still have drawbacks. 
The dynamic approaches are very time-consuming~\cite{Xin2017IdentifyingTP, Ye2021AutomatedPA} while the static approaches are more efficient but could be less precise~\cite{Wang2020AutomatedPC}. 
\revised{Additionally, building certain static systems could be hard if they require crafting specialized code patterns or features.}
\revised{In this study, we aim to advance static APCA approaches.}

Numerous static APCA approaches have been proposed in recent years. 
\revised{For example, Ye et al.~\cite{Ye2022AutomatedCO} introduced ODS, which identifies overfitting patches by utilizing statically extracted code features at the Abstract Syntax Tree (AST) level from both the buggy code and patches generated by APR tools.}
Tian et al.~\cite{tian2020evaluating} leveraged advanced code representation learning techniques such as BERT~\cite{bert}, to extract source code embeddings for assessing patch correctness.
Recently, Tian et al. introduced Quatrain~\cite{quatrain}, which transforms the APCA task into a question-answering problem. 
Quatrain first learned the relationship between bug reports and patch descriptions.
Subsequently, it constructed a question-answer-based classifier to assess patch correctness.

Moreover, Lin et al.~\cite{cache} proposed Cache, a patch correctness assessment technique that learns a context-aware code change embedding, considering program structures. 
Cache achieved the state-of-the-art performance and even outperformed many dynamic approaches~\cite{xiong2018identifying, Wang2020AutomatedPC, Yang2017BetterTC, Xin2017IdentifyingTP, Pacheco2007RandoopFR}.

Static APCA approaches, such as ODS~\cite{Ye2022AutomatedCO}, Tian et al.'s approach~\cite{tian2020evaluating},  Quatrain~\cite{quatrain}, and Cache~\cite{cache}, directly extract features from patches and learn correct patterns from the labeled dataset.
In prior works, static APCA approaches were primarily evaluated using $k$-fold cross-validation, where patches generated by different APR tools are mixed and separated into ($k$-1):1 for training and testing. 
However, the cross-validation setting has a significant limitation: for patches generated by a new or unseen APR tool, users are implicitly required to manually label a significant portion of these patches (e.g., 90\% in 10-fold cross-validation) before inferring the remaining patches (e.g., 10\% in 10-fold cross-validation).
Suppose Bob is a practitioner eager to leverage an advanced new APR tool to fix a bug and also aims to utilize APCA approaches to filter out the overfitting patches generated by the new tool.
However, the 10-fold cross-validation process implicitly necessitates that Bob manually labels 90\% of the patches generated by the new APR tool. This significant manual effort may deter Bob from adopting APCA approaches.

Given the continuous and rapid emergence of new APR tools, our goal is to alleviate the burden on users by avoiding the need for manual labeling of the patches generated by these new tools before APCA approaches can predict their correctness.
Additionally, many patches generated by existing APR tools have already been manually checked for correctness~\cite{Wang2020AutomatedPC,tian2020evaluating}. Thus, we are motivated to ask a key question:

\begin{tcolorbox}[enhanced,width=\linewidth,drop fuzzy shadow southwest,
    colback=gray!10]
\textit{Is it feasible to utilize \textbf{labeled patches of existing APR tools} to predict the correctness of the patches generated by a \textbf{new/unseen} APR tool?}
\end{tcolorbox}

We first explore whether the state-of-the-art APCA approaches can predict the correctness of patches generated by an unseen APR tool when labeled patches of other APR tools are available in the training data.   
During our preliminary experiments, we observed that state-of-the-art APCA approaches such as Quatrain~\cite{quatrain} and Cache~\cite{cache} did not yield satisfactory results. 
This was possibly due to a lack of labeled patches from the new or unseen APR tool for training the model. 
\revised{Addressing this challenge is critical for further advancing the APCA task.}
We report the effectiveness of these state-of-the-art approaches in Section~\ref{subsection:rq1}.
\revised{
While the setting proposed above has been studied by Ye et al.~\cite{Ye2022AutomatedCO}, it's worth noting that Ye et al. did not utilize patches manually labeled by developers. In contrast, our experiments are conducted on datasets that have undergone manual checks. This ensures a potentially higher level of reliability in the findings presented in our work. Additionally, in this study, we introduce a novel large language model-based solution to effectively tackle the challenge, which significantly outperforms the ODS approach proposed by Ye et al.~\cite{Ye2022AutomatedCO}.
}

\begin{figure}[t]  
\includegraphics[width=\linewidth]{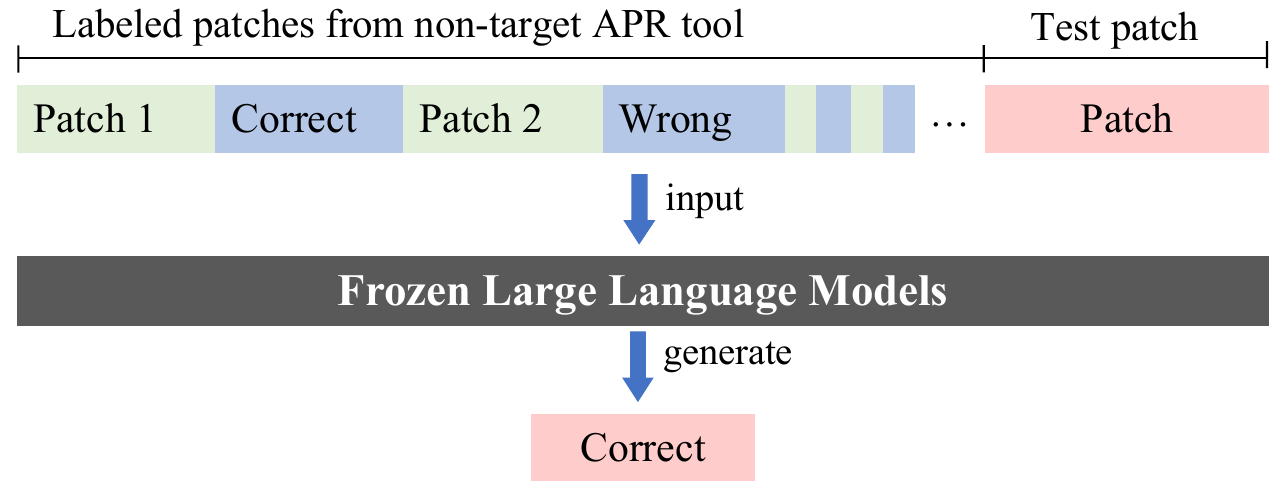} 
\vspace{-0.8cm}
\captionsetup{justification=raggedright}
\caption{An abstracted example of how to use the large pre-trained models. An unlabeled test patch of
a new APR tool is concatenated by several labeled patches of existing APR tools to form the input to the pre-trained model.}
\label{fig:simple_example} 
\end{figure}

\revised{To tackle the challenge, we propose \textbf{\toolname}, which aims to enhance the effectiveness of predicting the correctness of patches generated by new or unseen APR tools. 
\toolname employs an open-source large language model (LLM) for code, called Starcoder-7B~\cite{li2023starcoder}, to evaluate patch correctness, without requiring fine-tuning.
Technically, we directly leverage the pre-training objective of the LLM: generating the next token based on all previous tokens, to accomplish the APCA task.} 
As shown in Figure~\ref{fig:simple_example}, we first prepare model inputs where an unlabeled patch of a new APR tool is concatenated by several labeled patches of existing APR tools.
We then query the LLMs to generate the next token to show its tendencies in terms of patch correctness (i.e. generating a token either ``correct" or ``overfitting"). 
This allows us to apply LLMs without the need for fine-tuning since we formulate the APCA task (i.e. predicting whether a patch is correct or not) in the same format as the pre-training task (i.e., generating the next token).
The utilization of LLMs in this manner is referred to as \textit{In-context learning}.

\begin{figure*}[t] 
\centering  
\includegraphics[width=\linewidth]{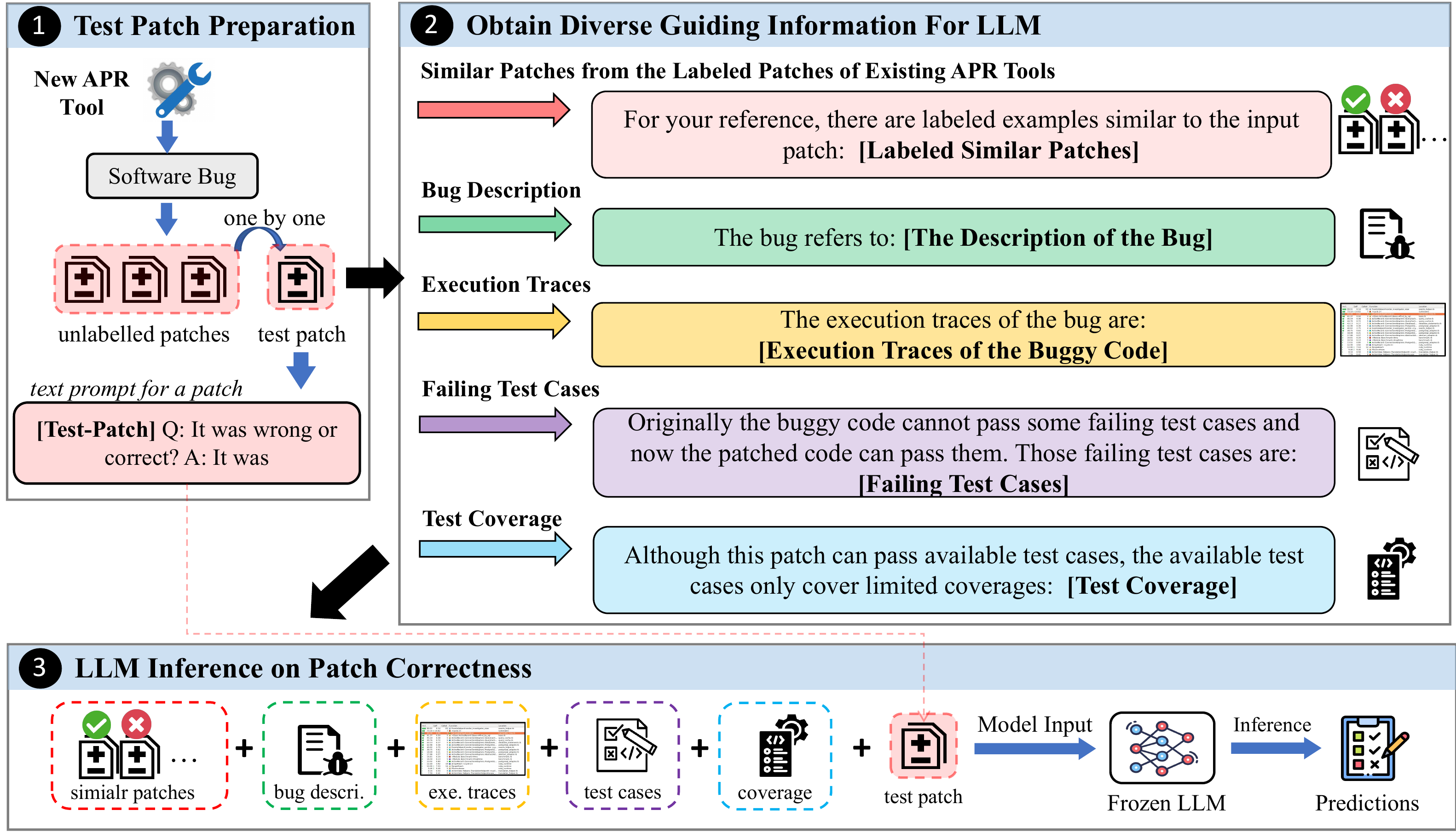} 
\centering 
\captionsetup{justification=centering}
\caption{Overall Framework of \toolname.}
\label{fig:framework} 
\vspace{-0.2cm}
\end{figure*}

\revised{Furthermore, \toolname does not encompass all labeled patches produced by existing APR tools; rather, it selects semantically similar patches. This strategy aids the large language model in providing more precise predictions for patches generated by a new APR tool.}
\revised{
In addition to the labeled patches of existing APR tools, LLM4PatchCorrect incorporates a broader range of guiding information, as illustrated in \textbf{\ding{183}} of Figure~\ref{fig:framework}:
\begin{enumerate}
    \item Bug Descriptions: descriptions detailing the nature of the bug that the patch intends to resolve;
    \item Execution Traces: traces of the buggy program's executions;
    \item Failing Test Cases: test cases that expose failures in the buggy program;
    \item Test Coverage: line and condition coverage metrics for all available test cases associated with the bug.
\end{enumerate}
Bug descriptions, execution traces, and failing test cases serve to enhance \toolname's comprehension of the characteristics of the bug targeted by a patch generated through a new APR tool. Test coverage serves as an approximate indicator of the adequacy of the available test cases. In cases where test coverage is notably low, even if a patch enables the program to pass all test cases, the correctness of the patch cannot be guaranteed because many code lines and conditions remain uncovered in the tests.
By leveraging the diverse range of guiding information (including the labeled patches of existing APR tools), \toolname can provide accurate predictions.
}

\revised{We evaluate \toolname using two real-world, large-scale patch correctness datasets~\cite{Wang2020AutomatedPC, tian2020evaluating}. These datasets comprise 1,179 patches generated by 22 different APR tools, with labels meticulously examined by developers.
The experimental results demonstrate that \toolname significantly improves Accuracy, F1, and AUC scores, increasing them from 10.2\% to 32.4\%, 6.1\% to 24.1\%, and 10.1\% to 33.2\%, on average, respectively, compared to state-of-the-art APCA approaches.}

\vspace{0.1cm}
\noindent\textbf{Contributions.} The main contributions are as follows:
\begin{itemize}[left= 6 pt, itemsep= 0.5 pt,topsep = 5 pt]
    \item This paper underscores the importance of a novel setting for the APCA task, where we assume that no labeled patches are available for a new or unseen APR tool. This setting can better match the initial goal of APCA tasks to reduce the manual labeling effort and can evaluate the ability of approaches to transfer knowledge embedded in the existing labeled data to future unlabeled data.

    \item \revised{To the best of our knowledge, we are the first to introduce advanced LLM in solving the APCA task. We design an LLM-based solution, LLM4PatchCorrect, for this challenging setting (i.e., no labeled patches of the new or unseen APR tool).}
    
    \item \revised{We propose incorporating diverse guiding information to aid \toolname in decision-making regarding patch correctness. Specifically, \toolname takes into account bug descriptions, execution traces, failing test cases, test coverage, and labeled patches generated by existing APR tools.}

\end{itemize}

\section{Background}
\label{sec:background}

\subsection{Large Language Models (LLMs) for Code}

Large Language Models (LLMs)~\cite{bert,roberta,CodeBERT,GraphCodeBERT,CodeT5,Codex,codeparrot,bloom} become popular in Natural Language Processing (NLP) and Software Engineering (SE). 
CodeBERT~\cite{bert} is a typical encoder-only model for code that is widely used in SE tasks such as code search and defect prediction. 
CodeT5~\cite{CodeT5} is a typical pre-trained encoder-decoder model for code, which is pre-trained on denoising sequence-to-sequence objectives.
Starcoder~\cite{li2023starcoder}, CodeLlama~\cite{codellama}, CodeParrot~\cite{codeparrot}, BLOOM~\cite{bloom} are typical LLMs that use only the Transformer decoder to predict the probability of the next token given the previous tokens. 
The nature of these models makes them highly useful for generation tasks because text/code is usually written in a left-to-right way. 

As shown in Table~\ref{table:model_diff}, except for the structure differences, these large language models for code are also different in model sizes, the maximum tokens can deal with, and the amount of pre-training data. Specifically, if we compare CodeBERT with Starcoder, CodeBERT can at most deal with a code snippet of 512 tokens while Starcoder can deal with a data instance at most consisting of 8,192 tokens. 
\revised{Besides, Starcoder is pre-trained on a giant pre-training data of over 1,000 billion tokens sourced from 80+ programming languages.}
However, CodeBERT is pre-trained on the CodeSearchNet~\cite{husain2019codesearchnet} dataset of 8 million code snippets or documentation including 6 programming languages.
For model sizes, the largest variant of Starcoder is about 120 times larger than CodeBERT. 

\begin{table}[b]
\caption{Open-sourced Large Language Models for Code}
\vspace{-0.2cm}
\large
\resizebox{\columnwidth}{!}{%
\begin{tabular}{l|c|c|c|c}
\hline
\textbf{Models}   & \multicolumn{1}{c|}{\textbf{Structure}} & \textbf{\begin{tabular}[c]{@{}c@{}}Model Size \\ (Billion)\end{tabular}} & \textbf{\begin{tabular}[c]{@{}c@{}}Max Length \\ (Token)\end{tabular}} & \textbf{\#Training Data} \\ \hline
\textbf{CodeBERT~\cite{CodeBERT}} & Encoder                                 & 0.13                                                                    & 512                                                                    & 8M instances             \\ \hline
\textbf{CodeT5~\cite{CodeT5}}   & Enc-Dec                         & 0.22                                                                     & 512                                                                    & 8M instances             \\ \hline
\textbf{CodeParrot~\cite{codeparrot}}    & Decoder                                 & 1.5                                                                       & 1,024                                                                  & 15B tokens                  \\ \hline
\textbf{BLOOM-1.1B~\cite{bloom}}    & Decoder                                 & 1.1                                                                      & 2,048                                                                  & 366B tokens              \\\hline
\textbf{BLOOM-3B~\cite{bloom}}    & Decoder                                 & 3.0                                                                      & 2,048                                                                  & 366B tokens              \\\hline
\textbf{BLOOM-7.1B~\cite{bloom}}    & Decoder                                 & 7.1                                                                      & 2,048                                                                  & 366B tokens              \\\hline
\textbf{BLOOM~\cite{bloom}}    & Decoder                                 & 176                                                                      & 2,048                                                                  & 366B tokens              \\ \hline
\textbf{Starcoder-1B~\cite{li2023starcoder}}    & Decoder                                 & 1.0                                                                      & 8,192                                                                  & 1,000B tokens             \\\hline
\textbf{Starcoder-3B~\cite{li2023starcoder}}    & Decoder                                 & 3.0                                                                      & 8,192                                                                  & 1,000B tokens             \\\hline
\rowcolor{LightCyan}
\textbf{Starcoder-7B~\cite{li2023starcoder}}    & Decoder                                 & 7.0                                                                      & 8,192                                                                  & 1,000B tokens             \\\hline
\textbf{Starcoder~\cite{li2023starcoder}}    & Decoder                                 & 15.5                                                                     & 8,192                                                                  & 1,000B tokens              \\\hline
\textbf{CodeLlama-7B~\cite{codellama}}    & Decoder                                 & 7.0                                                                     & 16,384                                                                  & 520B tokens              \\\hline
\textbf{CodeLlama-13B~\cite{codellama}}    & Decoder                                 & 13.0                                                                     & 16,384                                                                  & 520B tokens              \\\hline
\textbf{CodeLlama-34B~\cite{codellama}}    & Decoder                                 & 34.0                                                                     & 16,384                                                                  & 520B tokens              \\\hline
\textbf{CodeLlama-70B~\cite{codellama}}    & Decoder                                 & 70.0                                                                     & 16,384                                                                  & 520B tokens              \\\hline
\end{tabular}
}
\label{table:model_diff}
\end{table}

\vspace{0.1cm}
\noindent\textbf{\revised{Our LLM Choice.}}
\revised{We choose to use Starcoder-7B~\cite{li2023starcoder} in this work, as highlighted in Table~\ref{table:model_diff}. This selection is based on several considerations:}
\begin{enumerate}
    \item \revised{CodeBERT and CodeT5 are relatively small in model sizes.}
    \item \revised{CodeParrot and BLOOM have limitations on relatively short input lengths (1,024 and 2,048 tokens, respectively).} 
    \item \revised{CodeLlama series are pre-trained on less pretraining data compared to Starcoder series.}
    \item \revised{Larger models like Starcoder (15.5B) and CodeLlama (13B--70B) require extensive computational resources.}
\end{enumerate}
\revised{Starcoder-7B strikes a balance between model size and computational requirements, making it suitable for research purposes where computational resources may be limited compared to those available to tech companies. Additionally, its large input range (8,192 tokens) allows us to provide diverse guiding information, such as bug descriptions, execution traces, failing test cases, test coverage, and labeled patches generated by existing APR tools.}

\subsection{Usages of Large Language Models}
\subsubsection{Fine-tuning}
Fine-tuning a large language model for downstream tasks~\cite{bert, t5, CodeBERT, CodeT5} is a prevalent paradigm in the NLP and SE field. Fine-tuning utilizes the knowledge in language models to achieve better model initialization. Using a pre-trained language model for initialization often produces better results with enough labeled data. To adapt the language models to downstream tasks, fine-tuning trains the model in a supervised way. Specifically, given a dataset that consists of task-specific samples $X$ and corresponding labels $Y$, fine-tuning aims to find a set of parameters $\theta$ for the large language model so that $\theta = \mathop{\arg\max}_{\theta} P(Y|X, \theta)$.

\subsubsection{In-context Learning}
Though very effective and easy-to-use, fine-tuning usually requires relatively large labeled downstream task datasets to fine-tune \textit{all parameters} in large language models~\cite{codexglue}. Besides, fine-tuning demands large GPU resources to load and update all parameters of large language models~\cite{Hu2022LoRALA}.

An alternative popular approach proposed in GPT3 is in-context learning (ICL)~\cite{gpt3}, which induces a model to perform a downstream task by inputting examples to the model without any parameter update or training. 
ICL requires no gradient-based training and therefore allows a single model to immediately perform evaluations on different datasets. 
ICL mainly relies on the capabilities and knowledge that a large language model learned during its pre-training. 
The in-context learning makes predictions based on the probability of generating the next token $y$ given the unlabeled data instance $x$ and the context $C$, which includes $k$ labeled examples.
ICL outputs the token $y$ with the highest probability as the prediction for the unlabeled input data $x$. 
It can be expressed as: $y=\mathop{\arg\max}_{y} P_{PTM} (y |C, x),$
where $PTM$ denotes the large language model. $C$ is a context/demonstration created by concatenating k instances along with their corresponding labels i.e., $C = {x_1, y_1, x_2, y_2, ..., x_k, y_k}$. 
As shown in the illustration of Figure~\ref{fig:simple_example}, ICL asks the large language model to predict the correctness of a test patch given several labeled patches as the context, and the large language model outputs ``correct" because its probability as the next token is the highest.

\section{Proposed Approach}
\label{sec:model}

\toolname is proposed to utilize labeled patches of existing APR tools to predict the correctness of the patches generated by a new/unseen APR tool.
Hereafter, we denote a new or unseen APR tool as the \textit{\textbf{target APR tool}}.
\revised{The framework of \toolname is presented in Figure~\ref{fig:framework}.
It accepts a patch generated by a target APR tool as the raw input (depicted in \textbf{\ding{182}} of Figure~\ref{fig:framework}) and produces its correctness label (i.e., clean or overfitting).
Specifically, \toolname consists of the following four main steps:}

\begin{itemize}[leftmargin=*]
    \item \revised{\textbf{Step 1: Prepare Test Patches}  (\ding{182} of Figure~\ref{fig:framework}).} \revised{Initially, we gather the patches generated by the target APR tool to form the test set. For each patch within this test set, we append a text prompt to it. Subsequently, we convert the "text prompt + patch" combination into sub-tokens using the tokenizers of LLMs.}
     
     \item \revised{\textbf{Step 2: Obtain Similar Patches From Training Set} (\ding{183} of Figure~\ref{fig:framework}).}
     \revised{For each test patch in the test set, we utilize a contrastive learning-based retrieval module to retrieve several patches with high semantic similarity from the training set for each test patch.
     }

     \item \revised{\textbf{Step 3: Obtain Other Guiding Information} (\ding{183} of Figure~\ref{fig:framework}).}
     \revised{For each test patch in the test set, we extract the bug ID targeted by the test patch. We then query the bug benchmark to acquire relevant information about the bug, including bug descriptions, execution traces, failing test cases, and test coverage.
     }

    \item \revised{\textbf{Step 4: LLM Inference} (\ding{184} of Figure~\ref{fig:framework}).}
    \revised{We feed the test patch with all the guiding information into the LLM for code (Starcoder-7B). 
    The LLM then predicts the next token conditioning on the input. The predicted next token is then mapped to the patch correctness label. }  
\end{itemize}

\subsection{Test Patch Preparation}\label{subsec:preprocess}

\noindent\textbf{Forming Training/Test Sets.}
\revised{
In the upper part of \ding{182} of Figure~\ref{fig:framework}, when employing a new and advanced APR tool to address a detected software bug, it may generate many candidate patches capable of passing all available test cases for the identified bug. Subsequently, developers must determine which candidate patch is truly correct and should be implemented.
In this study, our goal is to utilize labeled patches from existing APR tools instead of asking developers to manually label those generated by a new or unseen APR tool. 
In line with this objective, we designate the labeled patches from existing APR tools as the \textit{\textbf{training set}}, and we designate the patches generated by new or unseen APR tools as the \textit{\textbf{test set}}. Moreover, we refer to the new or unseen APR tool as the \textit{\textbf{target APR tool}}.}

\vspace{0.1cm}
\noindent\textbf{Prompting Test Patches.}  
For each test patch, we pre-process it based on the recent advances in NLP, namely prompting\revised{~\cite{gpt3}},  which can help to better adapt a generic pre-trained model to a specific downstream task.
The intuition of prompting is to convert the downstream tasks into a similar form as the pre-training stage. 
For pre-trained models whose pre-training objective is to predict the next token given previous tokens, e.g., GPT-3~\cite{gpt3} and Starcoder~\cite{li2023starcoder}, prompting aims to ask a model to predict the next token (i.e. ``correct" or ``wrong" in this task) given previous tokens (patch contents and demonstrations).
To help large language models understand task-specific information, prompting modifies the input data by adding a piece of text description, namely prompt templates. 
\revised{We utilize the following prompt template for the test patches:}
\vspace{-0.1cm}
\begin{center}
\begin{tcolorbox}[colback={blue!5!white},
                  colframe=black,
                  width=8.5cm,
                  arc=1mm, auto outer arc,
                  boxrule=0.2pt,
                  top=0pt,
                  bottom=0pt
                 ]

\revised{$\{test\text{-}patch\}$ Q: It was wrong or correct? A: It was}
\end{tcolorbox}
\end{center}
\vspace{-0.1cm}
\revised{The $\{test\text{-}patch\}$ placeholder is replaced with the test patch content. The large language model (LLM) receives the test patch after patching as input and predicts the next token, indicating whether the patch is ``correct'' or ``wrong''. The process is also presented in \ding{182} of Figure~\ref{fig:framework}).} 

\begin{figure*}[t] 
\centering  
\includegraphics[width=0.9\linewidth]{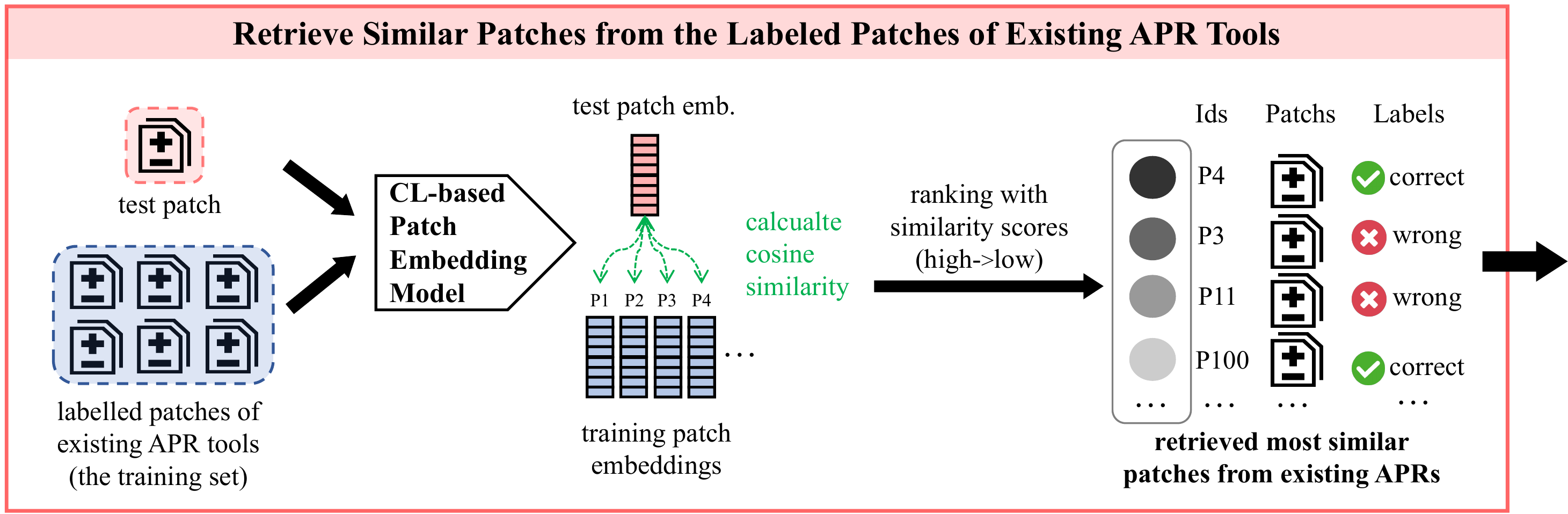} 
\centering 
\captionsetup{justification=centering}
\caption{The process of obtaining similar patches from the training set.}
\label{fig:framework_retrieval} 
\vspace{-0.2cm}
\end{figure*}

\revised{Please note that a generated patch is ``clean'' (``correct'') if it not only passes all the available test cases but also fixes the bugs in the program.
A generated patch is considered “overfitting” (``wrong'') if it only passes all the available test cases but is still incorrect with respect to the intended program specification which is not suitable for program repair.
Because the term ``overfitting'' (or ``clean'') has unique meanings in the APCA task, referring to an incorrect (or correct) patch, its interpretation can significantly differ in general English texts. 
To avoid confusion for the LLMs, we choose more commonly understood terms in both textual and code contexts, such as ``correct" and ``wrong", in the prompt.}

\vspace{0.1cm}
\noindent\textbf{Tokenization.}
To tokenize the inputs for the large language models, we use their corresponding tokenizers.
They are generally built based on the byte pair encoding (BPE)~\cite{BPE}, which outputs a sequence of sub-token sequences. 
BPE can reduce the size of the vocabulary by breaking uncommon long tokens into sub-tokens that frequently appear in the pre-training corpus. 
Besides, BPE is known to help mitigate the Out-of-Vocabulary (OoV) issue~\cite{Radford2019LanguageMA}.

\subsection{\revised{Obtaining Similar Patches From Training Set}}\label{subsec:demon}

\subsubsection{Motivation.}
\revised{Fine-tuning a large language model such as Starcoder-7B demands extensive computational resources. Hence, we opt to employ In-context learning (ICL) to adapt the LLM to the APCA task.} A standard ICL approach~\cite{gpt3} treats all the labeled training patches equally and randomly samples $k$ labeled patches to form a demonstration for the LLM, providing the context information about a downstream task. 
In the APCA task, however, every patch does not equally contribute. 
For instance, a patch that aims to fix a similar bug to the test patch is more instructive than a randomly sampled one. 
In other words, we need to design a retrieval module that can select the most valuable labeled patches for each test patch and inspire the LLM to achieve better prediction results.

To select patches, a simple-yet-effective idea is to choose semantically similar patches from the labeled training patches. The semantically similar patches contain similar code changes to the test patch. The difference between similar patches and test patches can possibly contribute to the difference in labels.
We assume that providing such semantically similar patch-label pairs to the LLMs can inspire them to learn the context information related to the test patch.

\subsubsection{Approach.}
\revised{The process of obtaining similar patches from the training set is illustrated in Figure~\ref{fig:framework_retrieval}. It involves the following four steps:}
\begin{enumerate}
    \item We first embed patches in both the training and test set into vector representations by utilizing a patch embedding model based on contrastive learning.
    \item For each test patch $x$, we retrieve its top-k most similar patches (i.e., $x_1, x_2, ..., x_k$) among the labeled training patches measuring the distances in the vector space by calculating cosine similarity.
    \item The top-k most semantically similar patches are modified via prompting (Section~\ref{subsec:preprocess}) and then concatenated to form the context/demonstration. 
    \revised{The top portion of Figure~\ref{fig:prompt} illustrates an example of the demonstration comprising similar patches from the training set.}
    \item  \revised{The demonstration, along with other guiding information (introduced in Section~\ref{subsec:demon_other}), is subsequently appended to the test patch $x$ and ultimately fed into the LLM.}
\end{enumerate}

\revised{Next, we introduce the construction of the contrastive learning-based patch embedding model, which serves as the core of the approach design and is utilized in Step 1 above.}

\subsubsection{Contrastive Learning-based Embedding Model.}
It is of great significance to represent patches in suitable embeddings because it will particularly affect whether the patches with the highest similarity scores are semantically similar to the test patch or not. 
To embed patches into embeddings of good quality, we train an unsupervised patch representation model by leveraging contrastive learning~\cite{SIMCSE}.

\begin{algorithm}[htb]
\caption{ Triplet data construction.}
\label{alg:Framwork}
\begin{algorithmic}[1] 
\REQUIRE ~~\\ 
    Patches of the ManySStuBs4J dataset, $P=[p_0, ..., p_n]$;\\
    The CodeBERT model, $CodeBERT(\cdot)$;\\
    The first dropout module, $Dropout_{1}(\cdot)$;\\
\ENSURE ~~\\ 
    Embeddings of Constructed triplets, $T=[t_0, ..., t_n]$ where $t_i$ is the embedding of a triplet $\langle p_i, p_{i}^{+}, p_{i}^{-}\rangle$ and
    $t_i = \langle emb(p_i), emb(p_{i}^{+}), emb(p_{i}^{-}) \rangle$ \\
    \FOR{each $i \in [0,n]$}
            \STATE obtain the embedding of $p_i$:\\
            $emb(p_i) = Dropout_{1}(CodeBERT(p_i))$;\\
            \STATE sample a different patch $p_j$ as $p_{i}^{-}$ where $i \neq j$;\ 
            \STATE obtain the embedding of $p_i^{-}$:\\
            $emb(p_i^{-}) = Dropout_{1}(CodeBERT(p_j))$;\\
            \STATE $emb(p_{i}^{+})$ = GetEmbeddingPositive ($p_i$)
            \STATE $t_i = \langle emb(p_i), emb(p_{i}^{+}), emb(p_{i}^{-}) \rangle$
            \STATE append $t_{i}$ to the list $T$\\
    \ENDFOR
\RETURN $T$; 
\end{algorithmic}
\end{algorithm}

\begin{algorithm}[htb]
\caption{ GetEmbeddingPositive: get the embedding of the positive sample $p_i^{+}$.}
\label{alg:getpositive}
\begin{algorithmic}[1] 
\REQUIRE ~~\\ 
    The input patch, $p_i$;\\
    The CodeBERT model, $CodeBERT(\cdot)$;\\
    The second dropout module, $Dropout_{2}(\cdot)$;\\
 
\ENSURE ~~\\ 
    Embedding of $p_i^{+}$, $emb(p_i^{+})$;\\
    \STATE use the second dropout module and CodeBERT to get the embedding of $p_i$:\\
    $emb_2(p_i) = Dropout_{2}(CodeBERT(p_i))$;\\
    \STATE regard the $emb_2(p_i)$ as $emb(p_i^{+})$:\\
    $emb(p_i^{+})=emb_2(p_i)$
\RETURN $emb(p_i^{+})$; 
\end{algorithmic}
\end{algorithm}

\vspace{0.1cm}
\noindent\textbf{\revised{Triplet Data Construction.}}
In the contrastive learning framework, we need a pre-training dataset whose input instance is in the form of a triplet $\langle p, p^{+}, p^{-}\rangle$,  where $p$ is a patch, $p^{+}$ is a semantically similar patch to $p$, and $p^{-}$ is a semantically dissimilar patch to $p$. 
Thus, $\langle p, p^{+} \rangle$ is considered as a similar pair and $\langle p, p^{-}\rangle$ is a dissimilar pair.
The main training objective of contrastive representation learning is to learn such an embedding space in which similar patch pairs stay closer to each other while dissimilar ones push out far away from each other.

\revised{As shown in Algorithm~\ref{alg:Framwork}, we construct the embeddings of triplets $\langle p, p^{+}, p^{-}\rangle$ from the ManySStuBs4J~\cite{manys} dataset, which contains 153,652 single-statement bug-fix patches mined from 1,000 popular open-source Java projects. This dataset is widely used in many SE downstream tasks such as automated program repair~\cite{mashhadi2021applying}, bug detection~\cite{Allamanis2021SelfSupervisedBD}, and fault localization~\cite{Wang2021BeepFF}.
For each patch $p$ in the ManySStuBs4J dataset, we perform the following steps to generate the embedding of a triplet $t_i = \langle emb(p_i), emb(p_{i}^{+}), emb(p_{i}^{-}) \rangle$:}
\begin{itemize}
    \item \revised{$emb(p_i)$: get one patch $p_i$ from the ManySStuBs4J dataset and feed $p_i$ to the CodeBERT and the first dropout module to get the embedding;}
    \item \revised{$emb(p_{i}^{-})$: sample another patch $p_j$ distinct from $p_i$ from the ManySStuBs4J dataset and feed $p_j$ to the CodeBERT and the first dropout module;}
    \item \revised{$emb(p_{i}^{+})$: obtain the embedding of the patch resulting from a semantic-preserving transformation applied to $p_i$ (Algorithm~\ref{alg:getpositive}).
    }
\end{itemize}
\revised{Patches $p_i$ and $p_i^{-}$ can be easily obtained from the ManySStuBs4J dataset, allowing us to easily obtain their embeddings as well. Thus, our primary concern lies in obtaining the embedding of $p_i^{+}$ (Algorithm~\ref{alg:getpositive}). It's crucial to note the \textbf{two criteria for a valid positive sample $p_i^{+}$}: 
1) The embedding of $p_i^{+}$ must differ from that of $p_i$ (i.e., $emb(p_i) \neq emb(p_i^{+})$) since the two patches are not identical after the transformation.
2) Despite this difference, the embeddings of $p_i^{+}$ and $p_i$ should encapsulate the same semantics.}

\begin{figure}[t] 
\centering 
\includegraphics[width=\linewidth]{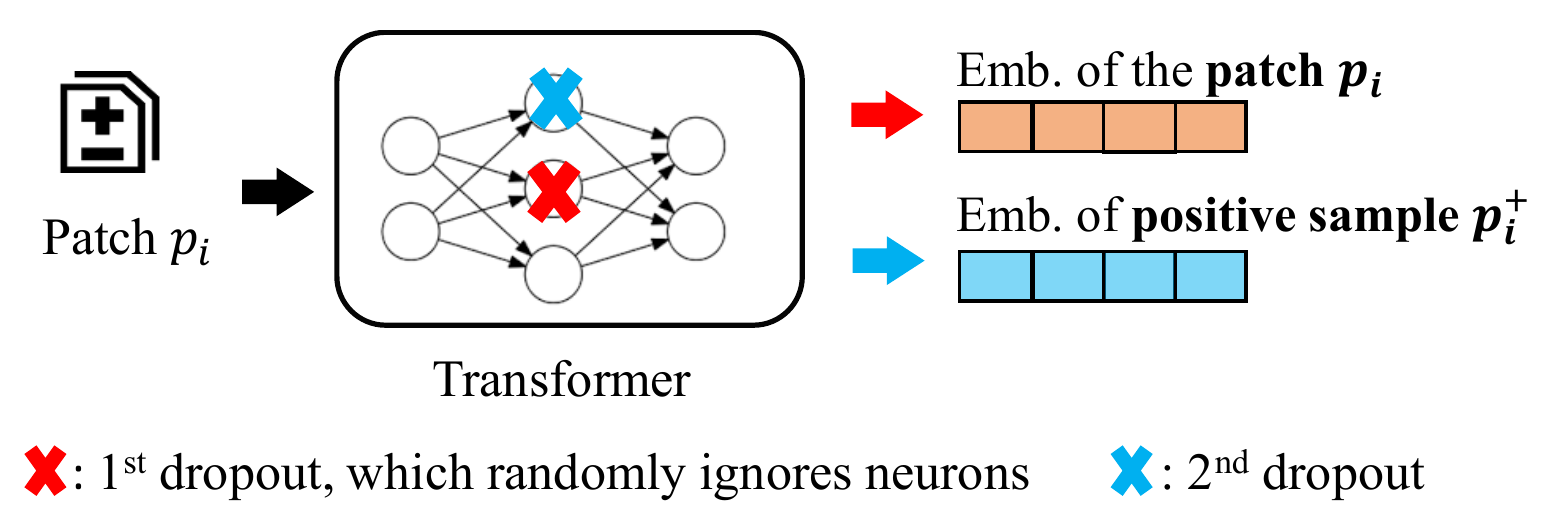} 
\captionsetup{justification=raggedright}
\vspace{-0.2cm}
\caption{Generating the embeddings of positive samples via the Dropout operation in SIMCSE. It adopts two different dropout operations which results in two different embeddings.}
\label{fig:dropout} 
\end{figure}

\begin{figure}[t] 
\centering 
\vspace{-0.2cm}
\includegraphics[width=0.55\linewidth]{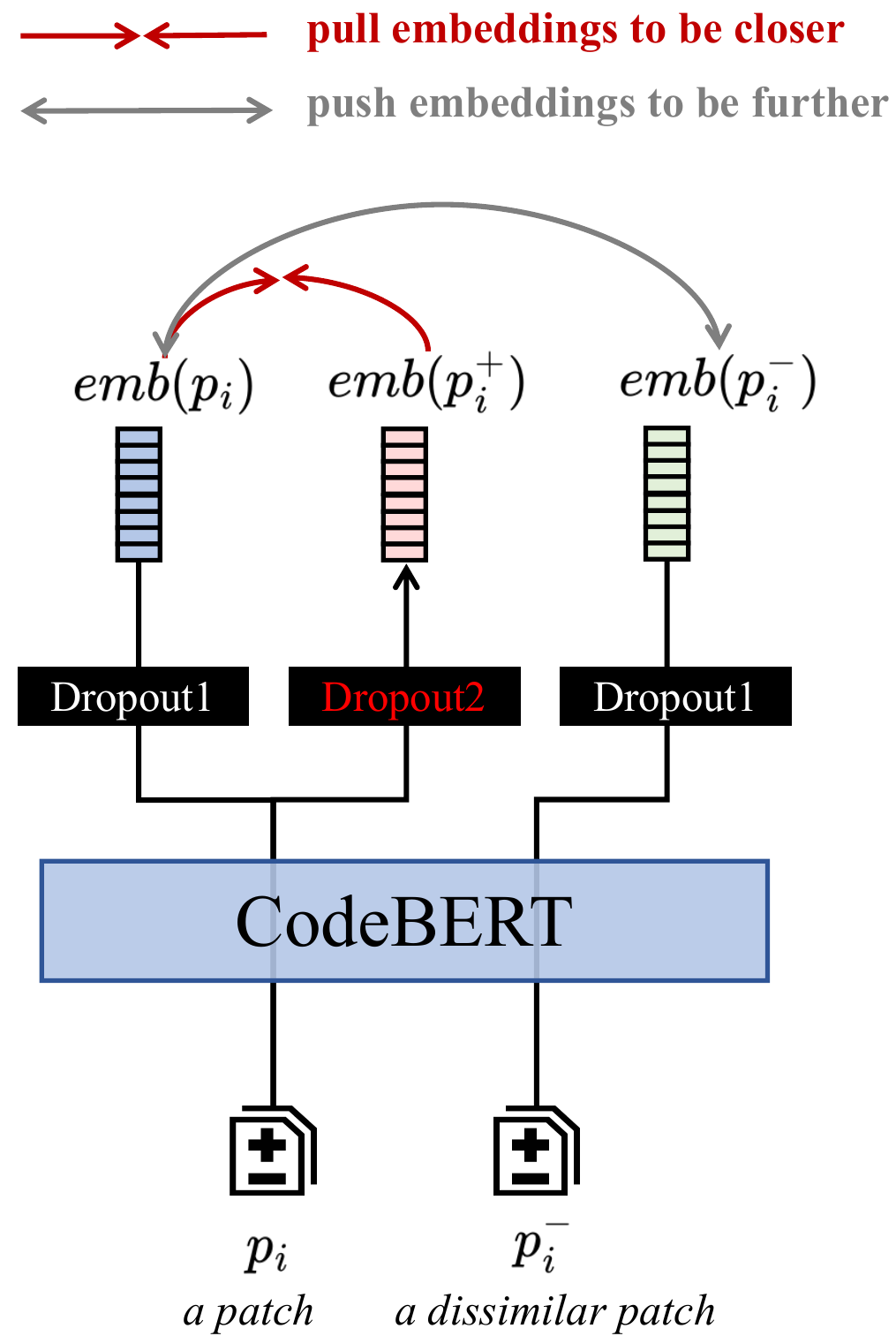} 
\vspace{-0.2cm}
\captionsetup{justification=raggedright}
\caption{Training CL-based patch embedding model with the embeddings of the constructed triplet $\langle p, p^{+}, p^{-}\rangle$.}
\label{fig:training_cl} 
\end{figure}

\begin{figure*}[t] 
\centering  
\includegraphics[width=0.85\linewidth]{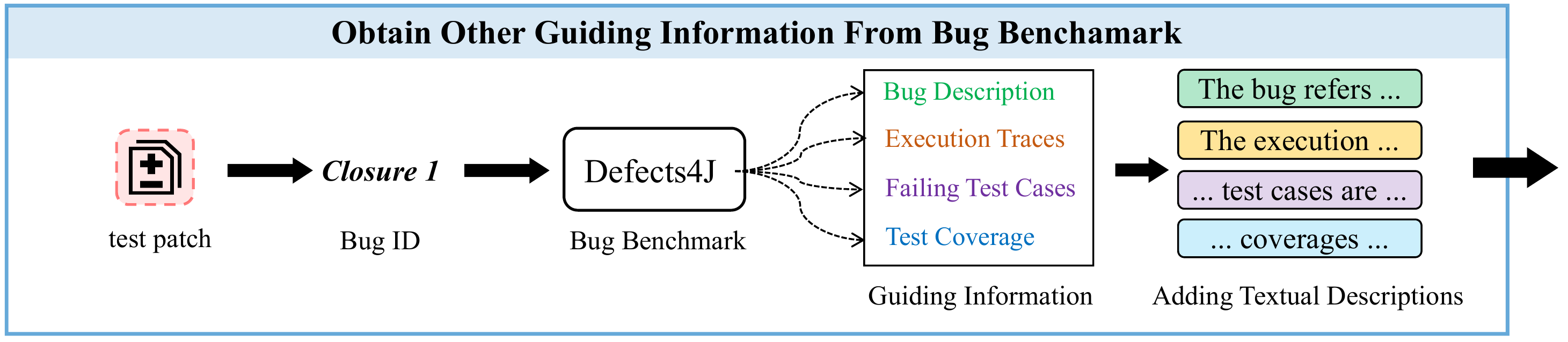} 
\centering 
\captionsetup{justification=centering}
\vspace{-0.2cm}
\caption{The process of obtaining the other guiding information from bug benchmark.}
\label{fig:framework_other_info} 
\vspace{-0.2cm}
\end{figure*}

To obtain the embedding of the positive sample $p^{+}$, we apply a semantic-preserving transformation proposed in SIMCSE~\cite{SIMCSE} on patches, which is based on the dropout operation. Dropout ~\cite{dropout} is a popular technique where randomly selected neurons are ignored during training to alleviate the overfitting problem of neural networks. 
\revised{As shown in Figure~\ref{fig:dropout}, SIMCSE simply applies the standard dropout operation~\cite{dropout} twice to obtain two different embeddings of the same patch $p_i$.
The \emph{first} dropout operation (the red one) randomly ignores a set of neurons, resulting in the embedding of patch $p_i$, denoted as $emb(p_i)$. Then, SIMCSE adopts the \emph{second} dropout operation (the blue one) which ignores \textit{another set of neurons}, also resulting in a different embedding $emb_2(p_i)$.}

\revised{The distinction between  $emb(p_i)$ and $emb_2(p_i)$ arises from the application of different dropout operations, satisfying the first criterion for a valid $p_{i}^{+}$.
Remarkably, despite this difference, both $emb(p_i)$ and $emb_2(p_i)$ originate from the same patch $p_i$ and inherently retain the same semantics, thereby fulfilling the second criterion for a valid $p_{i}^{+}$.}
 
\revised{Notably, $emb_2(p_i)$ meets the two criteria required for the embedding of a valid positive sample.
Consequently, the authors of SIMCSE regard $emb_2(p_i)$ as the approximation of the embedding of $p_{i}^{+}$ and simply utilize $emb_2(p_i)$ as $emb(p_{i}^{+})$, as shown in Algorithm~\ref{alg:getpositive}.
In essence, SIMCSE does not directly generate a positive patch $p_{i}^{+}$ from $p_i$; rather, it directly generates the embedding of the patch $p_{i}^{+}$ ($emb(p_{i}^{+})=emb_2(p_{i})$) through distinct dropout operations.
In other words, SIMCSE applies a transformation at the embedding level, modifying the embeddings, rather than altering the raw data (the patch) itself.}

\vspace{0.2cm}
\noindent\textbf{\revised{Training Details.}}
\revised{After building a large number of triplets from the ManySStuBs4J dataset, we then use those triplets to perform the training of the ContrastiveLearning (CL) based patch embedding model. The training details are presented in Figure~\ref{fig:training_cl}.}

Similar to SIMCSE~\cite{SIMCSE}, we also use a pre-trained Transformer model as the base model for generating embeddings and add a multi-layer perceptron (i.e., MLP) on top of it. 
SIMCSE uses RoBERTa~\cite{roberta} as the base model; however, as our data is source code, we use CodeBERT~\cite{CodeBERT}, which is pre-trained on both source code and texts.
Note that in the framework of contrastive learning, fine-tuning is indispensable to learn such an embedding space in which similar patch pairs stay closer to each other. However, fine-tuning a huge LLM like Starcoder requires vast computation resources that we cannot afford. Thus, we choose to use CodeBERT to learn the contrastive learning-based patch embedding rather than the Starcoder.

\revised{
As depicted in Figure~\ref{fig:training_cl}, a patch $p_i$ and another distinct patch $p_i^{-}$ from the ManySStuBs4J dataset serve as inputs to the CL-based model during training. We employ the first dropout module (and the CodeBERT) to obtain their respective embeddings $emb(p_i)$ and $emb(p_i^{-})$. Following this, we utilize a second dropout module, distinct from the first one, on the patch $p_i$ to acquire the embedding $emb(p_i^{+})$.
Then, we calculate the loss using the following formula:
}
\begin{equation}
    l_i =  -log \frac{e^{ \sfrac{cos(emb(p_i), emb(p_i^{+}))}{\tau}}}{ \sum_{j}^{N} (e^{\sfrac{cos(emb(p_i), emb(p_j^{+})}{\tau}} + e^{\sfrac{cos(emb(p_i), emb(p_j^{-}))}{\tau}} ) }
\end{equation}
\revised{
where $p_i$ denotes the $i$-th patch and $p_j$ refers to the $j$-th patch which iterates over all patches in the mini-batch.
$N$ is the
number of patches in a mini-batch; $\tau$ is a temperature hyper-parameter; and 
$cos$ is the cosine similarity function.
By minimizing the loss above, the model learns such an embedding space in which similar patch pairs stay closer to each other while dissimilar ones push out far away from each other.
Then, we can utilize the trained CL-based model to assist in identifying similar patches from the training set when presented with a test patch.}
For other implementation details, we implement the model by using a popular deep-learning library called HuggingFace\footnote{https://huggingface.co/}. 
We simply adopt the hyper-parameters recommended by SIMCSE~\cite{SIMCSE} which are the learning rate as 5e-5, batch size as 64, and the number of epochs as 3.

\subsection{\revised{Obtaining Other Guiding Information}}\label{subsec:demon_other}

\subsubsection{Motivation.}
\revised{
In addition to the labeled patches of existing APR tools (the training set), LLM4PatchCorrect incorporates a broader range of guiding information, as illustrated in Figure~\ref{fig:framework_other_info}. 
Bug descriptions, execution traces, and failing test cases contribute to \toolname's understanding of the bug characteristics addressed by a patch generated through a new APR tool. Test coverage acts as an approximate gauge of the adequacy of the available test cases. In instances where test coverage is notably low, the correctness of a patch cannot be reliably ensured, even if the patch enables the program to pass all test cases, as numerous code lines and conditions remain uncovered.
}

\subsubsection{Approach.}
\revised{The process of obtaining the additional guiding information is illustrated in Figure~\ref{fig:framework_other_info}.
Firstly, when given a test patch generated by a new APR tool, we will inspect the metadata of the test patch to determine the ID of the bug that the new APR tool is addressing.
Secondly, we will use the bug ID to gather the following guiding information:
}
\begin{enumerate}
    \item \revised{Bug Descriptions: Descriptions detailing the nature of the bug that the patch intends to resolve;}
    \item \revised{Execution Traces: Traces of the buggy program's executions;}
    \item \revised{Failing Test Cases: Test cases that expose failures in the buggy program;}
    \item \revised{Test Coverage: Line and condition coverage metrics for all available test cases associated with the bug.}
\end{enumerate}
\revised{Lastly, we include the textual description, as depicted in \textbf{\ding{183}} of Figure~\ref{fig:framework}, to each piece of information. For instance, for the bug description, we use:}
\vspace{-0.1cm}
\begin{center}
\begin{tcolorbox}[colback={blue!5!white},
                  colframe=black,
                  width=8.5cm,
                  arc=1mm, auto outer arc,
                  boxrule=0.2pt,
                  top=0pt,
                  bottom=0pt
                 ]

\revised{The bug refers to: [The Description of the Bug]
}
\end{tcolorbox}
\end{center}
\vspace{-0.1cm}
\revised{The placeholder [The Description of the Bug] is replaced with the bug description obtained from the bug benchmark. Similarly, we utilize the textual descriptions from \textbf{\ding{183}} of Figure~\ref{fig:framework} for bug descriptions, execution traces, failing test cases, and test coverage.}

\revised{For implementation details, we depend on a bug benchmark such as Defects4J \cite{defect4j} to gather the guiding information from the bug ID. For example, within the Defects4J benchmark, once properly installed, users can compile a bug, conduct testing, and compute testing coverage using the following straightforward code:
}
\begin{tcolorbox}[colback=gray!20]
\begin{verbatim}
#Check out the buggy version from bug-id
checkout -p Lang -v 1b -w Lang1  

cd Lang1           #visit the bug
defects4j compile  #compile
defects4j test     #run the test cases
defects4j coverage #compute coverages
\end{verbatim}
\end{tcolorbox}
\revised{Upon executing the code above, users can find bug information, execution traces, failing test cases, and the testing coverages (for line and conditions) in the Lang1 folder. Consequently, we have obtained the additional information for the test patch.}

\subsection{\revised{LLM Inference on Patch Correctness}} \label{subsec:inference}

\subsubsection{Combining Diverse Guiding Information}

\revised{
\toolname utilizes a wide range of guiding information. Before conducting the LLM inference, we concatenate each piece of information together and then append the test patch.
The concatenation is performed as follows: 
$$C = [S_1; S_2; ... ; Bug; Trace; Case; Coverage; Test\text{-}Patch]$$
where ``$;$'' indicates the concatenation operation. $S_j$ represents the $j$-th retrieved similar patch from the training set.
Please note that each piece of information is prompted with the corresponding textual description, as depicted in \textbf{\ding{183}} of Figure~\ref{fig:framework}.
}

\begin{figure}[t] 
\centering  
\includegraphics[width=\linewidth]{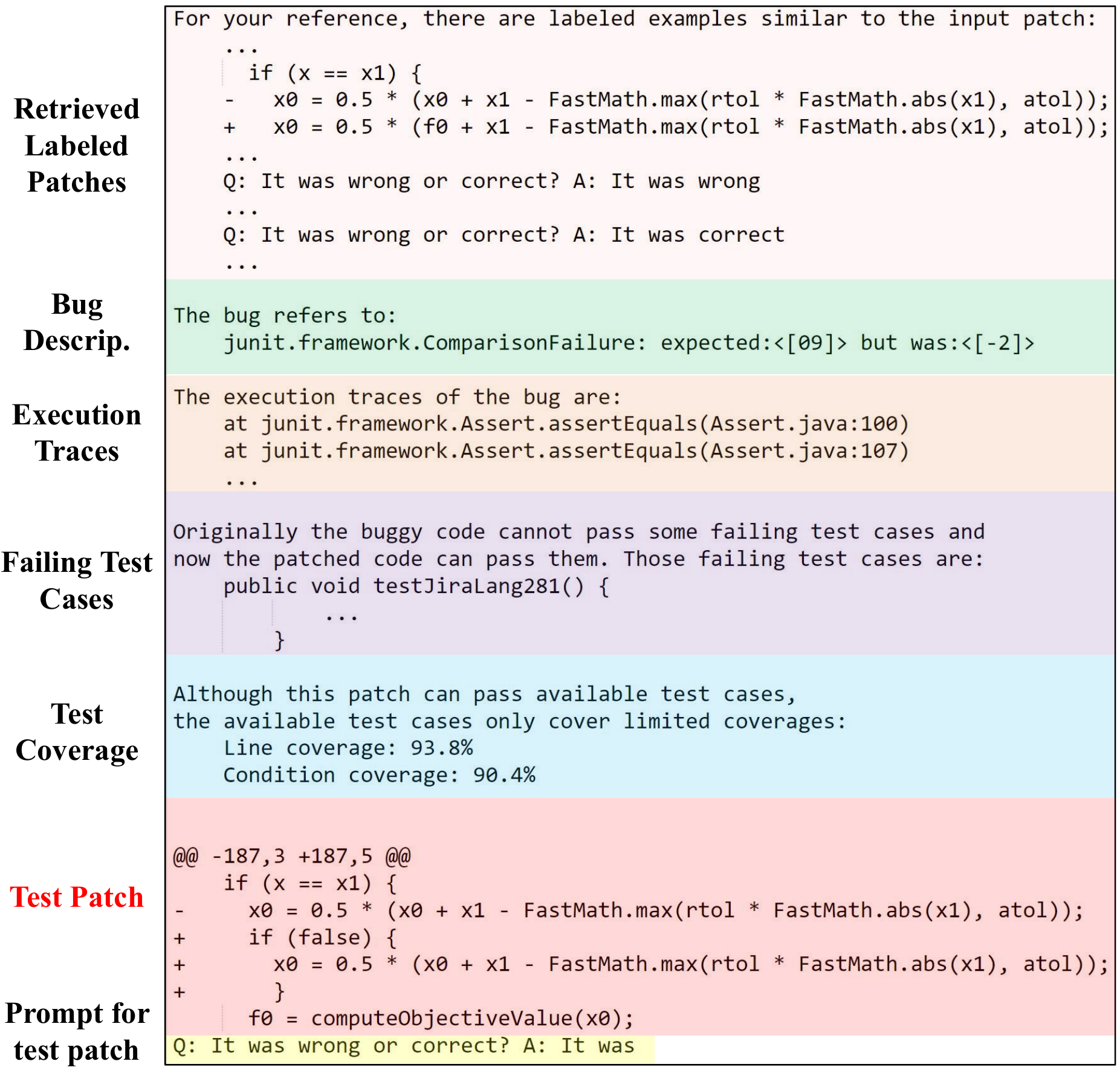} 
\centering 
\captionsetup{justification=centering}
\vspace{-0.2cm}
\caption{Example of the concatenated input to the LLM.}
\label{fig:prompt} 
\vspace{-0.2cm}
\end{figure}

\revised{
For better visualization, Figure~\ref{fig:prompt} presents one example of the final concatenated input (i.e., $C$) to the LLM, where the test patch is generated by DynaMoth and aims to address the Math-50 bug in Defects4J.}

\subsubsection{In-context Learning Inference}
The in-context learning inference can be regarded as a text generation problem where the LLM is frozen. 
Given the concatenated final input $C$ to the LLM, the in-context learning inference outputs the next token $y$ with the highest probability as the prediction for the unlabeled input data $C$. It can be formally expressed as: 
\begin{equation}
	y = \mathop{\arg\max}_{y} \ \ P_{LLM} (y|C).
\end{equation}
where LLM denotes the parameters of the pre-trained model which are frozen all the time. 
We use a recently proposed LLM, Starcoder-7B~\cite{li2023starcoder}, the multilingual LLM that is open-sourced, as the backbone model in \toolname.
Starcoder has a list of variants of different model sizes and the largest variant model has 15.5 billion parameters. 
\revised{With the help of the quantization~\cite{quantization} technique, which reduces memory and computational costs by representing weights and activations with lower-precision data types like 8-bit integers (int8), we are able to use any LLM within 7B parameters using a 12GB GPU (the GPU memory of a widely used GPU card 2080-Ti). This indicates we can use any LLM below 7B.
}

As the patch correctness assessment task is formulated as a binary classification task~\cite{tian2020evaluating, cache}, we first compute the probabilities of the clean class (i.e. $P_{PTM} (``correct" |C, x)$) and the overfitting class (i.e. $P_{PTM} (``wrong" |C, x)$) and normalize the probabilities of two classes. Then, if the probability of the overfitting class is larger than 0.5, the test patch is predicted as ``overfitting''. Otherwise, it is predicted as ``clean''.
We call the probability of the overfitting class generated by a large pre-trained model (after normalization) as the prediction score.
\revised{Please kindly note that a generated patch is ``clean'' (``correct'') if it not only passes all the available test cases but also fixes the bugs in the program.
A generated patch is considered “overfitting” (``wrong'') if it only passes all the available test cases but is still incorrect with respect to the intended program specification which is not suitable for program repair.
Because the term ``overfitting'' (or ``clean'') has unique meanings in the APCA task, referring to an incorrect (or correct) patch, its interpretation can significantly differ in general English texts. 
To avoid confusion for the LLMs, we choose more commonly understood terms in both textual and code contexts, such as ``correct" and ``wrong", when calculating the probabilities of the candidate next tokens.}

\section{Experimental Setting}
\label{sec:experiment}

\subsection{Dataset}
We use the dataset used in Lin et al.~\cite{cache}'s work containing a total of 1,179 patches from the Defects4J benchmark~\cite{defect4j} where most existing APR tools are evaluated. 
The dataset is merged from two existing large-scale datasets provided by Wang et al.~\cite{Wang2020AutomatedPC} and Tian et al.~\cite{tian2020evaluating}. 
Patches in these two datasets were either written by developers (i.e., the ground-truth patches) or generated by 22 different APR tools. Note that their correctness has been carefully labeled and checked.

\revised{To ensure the absence of duplicate patches generated by different APR tools, we conducted two verification steps. Firstly, we confirmed the absence of duplicated patches (i.e., identical patches) in the experiment dataset, employing string matching. Secondly, we verified the presence of semantic duplicates—patches that are not identical but possess semantic equivalence—through manual examination of all patches. In this process, we identified two pairs demonstrating semantic equivalence: 
1) a pair of patches generated by ACS for the Math-4 bug in Defects4J and 2)
a pair of patches generated by ACS for the Lang-35 bug in Defects4J.
We do not find other pairs of semantic equivalent patches in this process. To address the impact of the semantic equivalent patches, we have removed pairs of semantically duplicate patches from our evaluation data.
}

\subsection{Cross-Tool Validation}
In this paper, we aim to utilize labeled patches of existing APR tools to predict the correctness of the patches generated by a new/unseen APR tool.
However, patches generated by future APR tools are impossible to get. Thus, we conduct a ``cross-tool'' validation that is close to the setting above. 
In the ``cross-tool'' scenario, we iteratively regard each APR tool as the target APR tool. For instance, we can first consider the tool {\tt TBar}~\cite{Liu2019TBarRT}\ 
as the target APR tool, then, all the patches generated by {\tt TBar} are used as the test dataset. At the same time, other patches generated by other APR tools are used as the training data.
In this way, we can still evaluate whether the model can transfer the knowledge in
labeled patches of APR tools except {\tt TBar} to the patches generated by {\tt TBar}. 
To carry out the experiment, we employ 22 existing APR tools and construct 22 different sub-datasets in a leave-one-out manner, i.e., we iteratively pick one APR tool as the target APR tool for each sub-dataset.
Please note that we remove the patches of the existing APR tools (i.e., the labeled patch pool) that are identical to any patch in the test set to avoid the data leaking issue.

\subsection{Baselines}
Please note that each baseline tool is trained on the corresponding training set (i.e., the labeled patches generated by other APR tools) while considering each APR tool as the target at a time. 
Only \toolname does not have a training phase, while all the other methods need training.
Our baselines are listed as follows:

\vspace{0.1cm}
\noindent\textbf{\revised{Patch-Sim~\cite{xiong2018identifying}.}}
\revised{
Patch-Sim~\cite{xiong2018identifying} is a dynamic-based Automated Program Correctness Assessment (APCA) approach that relies on behavior similarities between program executions. It operates under the assumption that passing tests on original and patched programs are likely to behave similarly while failing tests on original and patched programs are likely to behave differently. Based on this observation, Patch-Sim generates new test inputs to enhance the test suites using Randoop. It then utilizes the behavior similarity of these test inputs to determine the correctness of patches.
}

\vspace{0.1cm}
\noindent\textbf{CodeBERT~\cite{CodeBERT}.} 
Following the success of transformer-based pre-trained models in NLP like BERT~\cite{bert}, researchers have proposed pre-trained models for code, e.g., CodeBERT~\cite{CodeBERT}. 
CodeBERT is a solid baseline in a wide range of SE downstream tasks such as code search and code summarization.
Fine-tuning models typically result in better effectiveness performance compared to freezing models~\cite{Sanh2022MultitaskPT}. Additionally, CodeBERT has only 0.13 billion parameters and can be fine-tuned using academic computational resources, such as one 2080-Ti GPU card.
Therefore, we directly fine-tune CodeBERT with the training data. We adopt the hyperparameters used in the CodeBERT paper~\cite{CodeBERT}: the learning rate is set to 1e-5 and the number of training epochs is 8.

\vspace{0.1cm}
\noindent\textbf{Tian et al.'s Approach~\cite{tian2020evaluating}.}
Tian et al.~\cite{tian2020evaluating} proposed a static-based APCA approach to leverage code (change) representation techniques to predict the correctness (i.e., correct or overfitting) of APR-generated patches. 
They adopted three recent representation techniques (i.e., BERT~\cite{bert}, CC2vec~\cite{cc2vec}, and Doc2vec~\cite{doc2vec}) with well-known Machine Learning classifiers (i.e., Logistic Regression, Decision Tree, and Naive Bayes) to demonstrate that it could achieve a promising result. 
Technically, they froze the representation models and used them to embed patches into distribution embeddings. Then, they fed the embeddings with the labels of patches to Machine Learning classifiers to train classifiers.

\vspace{0.1cm}
\noindent\textbf{\revised{ODS~\cite{Ye2022AutomatedCO}.}}
\revised{
Ye et al.~\cite{Ye2022AutomatedCO} propose ODS, a static-based approach for automated program correctness assessment, which leverages static code features.
ODS performs a comparison between a patched program and a buggy program to extract static code features at the abstract syntax tree (AST) level. Then, it employs the extracted code features along with patch correctness labels to train a learning-based model.
}

\vspace{0.1cm}
\noindent\textbf{\revised{Quatrain~\cite{quatrain}.}}
\revised{Tian et al.~\cite{quatrain} introduced Quatrain, a static-based approach in Automated Program Correctness Assessment (APCA), which redefines patch correctness evaluation as a question-answering task. Initially, Quatrain employs the natural language processing (NLP) technique to understand the relationship between bug reports and patch descriptions. Subsequently, it constructs a question-answer-based classifier to assess patch correctness.}

\vspace{0.1cm}
\noindent\textbf{Cache~\cite{cache}.} 
Lin et al.~\cite{cache} proposed an approach named Cache that showed the state-of-the-art performance in the patch correctness assessment task.
Cache learns a context-aware code change embedding considering program structures.
Specifically, given a patch, Cache focuses on both the changed code and correlated unchanged part and utilizes the AST paths technique for representation where the structure information from the AST node can be captured. 
After learning the representation, Cache builds a deep learning-based classifier to predict the correctness of the patch.
They performed extensive experiments and showed that Cache outperformed the existing representation learning-based techniques~\cite{tian2020evaluating} and testing-based approaches~\cite{xiong2018identifying, Wang2020AutomatedPC, Yang2017BetterTC, Xin2017IdentifyingTP, Pacheco2007RandoopFR}.

\subsection{Evaluation Metrics}
To evaluate the effectiveness of various target approaches, we adopt widely used evaluation metrics for classification tasks: Accuracy and F1-score.
Both Accuracy and F1-score can be measured based on the number of true positives (TP), false positives (FP), and false negatives (FN).
Accuracy is defined as the ratio of the number of correctly predicted data (i.e., TP+TN) to the number of all patches (i.e., TP+TN+FP+FN).
TP case is referred to when a model prediction is overfitting for an overfitting patch, otherwise, it is an FN case. 
FP case is referred to when a model prediction is overfitting for a correct patch, otherwise, it is a TN case. 
F1-score is defined as the harmonic mean of Precision and Recall values.
For example, Precision is the ratio of correctly predicted overfitting patches to all the patches predicted as overfitting (i.e., $ Precision = \frac{TP}{TP + FP} $) and Recall is the ratio of the number of correctly predicted overfitting patches to the actual number of overfitting patches (i.e. $ Recall   = \frac{TP}{TP + FN} $).
F1-score can be formally defined as $ F1\text{-}score = \frac{2 \times Precision \times Recall}{Precision + Recall} $.

\revised{
As cross-tool validation yields different results for different target APR tools, for easier comparison among methods, we will compute both ``averaged results'' and ``weighted averaged results'' across all the APR tools. In the "weighted averaged results", the weights are assigned based on the number of patches in the test sets. The formula for the weighted average is as follows:}
    $$weighted = \\
    \frac{\sum_{i=1}^{22} (\#\text{patch of the tool } i)\times (\text{metric of tool }i)}{\sum_{i=1}^{22} \#\text{patch of the tool } i}$$
\revised{
where ``tool $i$" refers to the target APR tool and ``$\#$ patch of the tool $i$" refers to the number of patches generated by the target APR tool $i$. 
``metric of tool $i$'' refers to the metric of a model for the target APR tool $i$, such as Accuracy and F1-scores. 
}

\revised{When calculating the improvement ratios on the (weighted) averaged metrics, we first compute the (weighted) averaged metric across all APR tools. Then, we determine the relative improvement ratios based on the (weighted) average results of the two models.
For example, if a patch dataset is generated by three APR tools and a baseline achieves F1-scores of 80\%, 70\%, and 60\% when considering each of the three APR tools as the target tool, and our model achieves F1-scores of 85\%, 75\%, and 65\% accordingly. To compute the relative improvement ratio on F1 based on the average results, we first compute the averaged F1-scores: 70\% (i.e., $\frac{80\%+70\%+60\%}{3}$) for the baseline, and 75\% (i.e., $\frac{85\%+75\%+65\%)}{3}$) for our model. Then, the relative improvement ratio of our model over the baseline will be 7.14\% (i.e., $\frac{75\%-70\%}{70\%} \times 100\%$).
}

\subsection{Hyper-parameter Tuning on Selecting Similar Labeled Patches}

\begin{figure}[t] 
\centering 
\includegraphics[width=\linewidth]{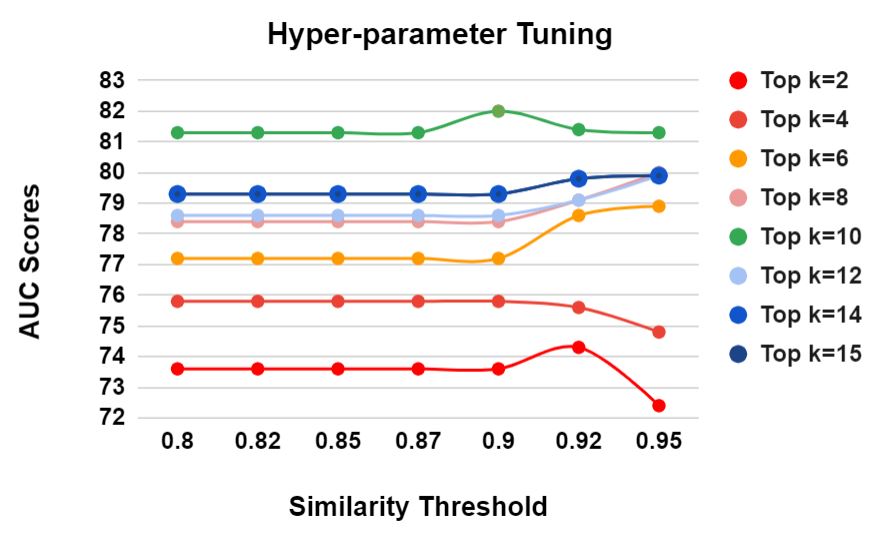} 
\captionsetup{justification=raggedright}
\caption{Hyper-parameter Tuning Results (AUC) of \toolname}
\label{fig:hp_result} 
\end{figure}

As introduced in Section~\ref{subsec:demon}, the final input to \toolname includes the most similar patches retrieved from the labeled patches of existing APR tools.
It is important to decide the way to choose the ``most similar patches'' given a test patch.
The general idea is to include as many similar patches as possible, such that it can provide more valuable information to \toolname.
We specify the general idea into two hyper-parameters:
\begin{itemize}
    \item the \textbf{$k$} value: it is the maximum number of the most similar patches that we consider to build the in-context learning demonstration. For instance, ``$k=10$'' indicates that we consider up to 10 patches.
    
    \item the similarity threshold \textbf{$\beta$}: it constrains the similarity of patches for building the demonstration. Specifically, for a test patch, only labeled patches with higher cosine similarities than the threshold \textbf{$\beta$} are considered.
\end{itemize}
\revised{To determine a fixed similarity threshold ($\beta$) and the maximum number of the most similar patches ($k$), we conducted preliminary experiments on a randomly split 5\% of the labeled patch pools. Specifically, we explored a fine-grained hyper-parameter range: $\beta$=[0.80, 0.82, 0.85, 0.87, 0.9, 0.92, 0.95] and $k$=[2, 4, 6, 8, 10, 12, 14, 15]. As illustrated in Figure~\ref{fig:hp_result}, \toolname exhibits the best performance when k=10 and beta=0.9.
}
We select the top 10 patches whose cosine similarity scores are higher than 0.9 to a test sample to form the final input to the model.

\subsection{Research Questions}
In this paper, we aim at answering the following research questions:
\begin{itemize}[leftmargin=*]
    \item \textbf{RQ1:} How does \toolname perform compared to state-of-the-art approaches in the cross-tool setting?
    \vspace{0.1cm}
    \item \textbf{RQ2:} How does each component of \toolname contribute?
    \vspace{0.1cm}
\end{itemize}

We design RQ1 to demonstrate the effectiveness of \toolname by comparing it with state-of-the-art patch correctness assessment approaches. In RQ2, we carefully conduct an ablation study to illustrate the contribution of each component in our solution.

\section{Experimental Results} 
\label{results}

\subsection{RQ1. Overall Performance of the Approach} 
\label{subsection:rq1}

\noindent\textbf{Main Results.}
We evaluate the effectiveness of \toolname by comparing it against the state-of-the-art APCA approaches. 
As we mentioned in the experimental setting, Tian et al.~\cite{tian2020evaluating} adopted three representation techniques and three machine learning classifiers.
Among nine variants, we only report the best-performing combination due to the limited space.
For Cache, we reuse the implementation released by the authors of Cache. 
Tables~\ref{table:acc}--\ref{table:auc} show the effectiveness of all the approaches including \toolname and the baselines in terms of Accuracy, F1-score, and AUC.

\revised{
The experimental results show that \toolname outperforms all the baseline techniques.
Specifically, \toolname showed \revised{20.9\% ($\frac{84.4-69.8}{69.8}$)}, \revised{13.1\% ($\frac{86.5-76.5}{76.5}$)}, and \revised{33.3\% ($\frac{80.4-60.3}{60.3}$)} improvements against Tian et al.'s work on average in terms of Accuracy, F1-score, and AUC. 
It also showed \revised{14.7\% ($\frac{84.4-73.6}{73.6}$)}, \revised{6.8\% ($\frac{86.5-81}{81}$)}, and \revised{30.7\% ($\frac{80.4-61.5}{64.5}$)} of enhancement against Cache on average in terms of Accuracy, F1-score, and AUC. 
Additionally, \toolname led to a substantial boost over a strong baseline CodeBERT by \revised{10.3\% ($\frac{84.4-76.5}{76.5}$)}, \revised{6.1\% ($\frac{86.5-81.5}{81.5}$)}, and \revised{10.1\% ($\frac{80.4-73}{73}$)} in Accuracy, F1-score, and AUC, respectively.
Besides, we adopt weighted averages of Accuracy and F1-score where the weights are the number of patches generated by each APR tool. 
\toolname still outperforms all baselines. 
\revised{For example, \toolname leads to 10.2\%, 4.7\%, and 26.5\% improvements over Cache in terms of weighted average Accuracy, F1-score, and AUC.
\toolname outperforms CodeBERT by 9.2\%, 5.0\%, and 6.1\% in terms of weighted average Accuracy, F1-score, and AUC.} 
}
Furthermore, we conduct the Wilcoxon signed-rank tests between \toolname and all baselines to investigate whether the improvements are significant. The results show that \toolname is statistically significantly better than all baselines (all p-values are less than 0.05). 

\begin{table}[]
\caption{Accuracy of \toolname and Baselines.}
\vspace{-0.2cm}
\huge
\label{table:acc}
\resizebox{\columnwidth}{!}{%
\begin{tabular}{l|b|r|r|r|r|r|g|c}
\hline
\textbf{\begin{tabular}[c]{@{}l@{}}Target \\ APR\end{tabular}}       & \textbf{\begin{tabular}[c]{@{}c@{}}correct:\\ wrong\end{tabular}} & \multicolumn{1}{c|}{\textbf{Cache}} & \multicolumn{1}{c|}{\textbf{Tian's}} & \multicolumn{1}{c|}{\textbf{CBERT}} & \multicolumn{1}{c|}{\textbf{ODS}} & \multicolumn{1}{c|}{\textbf{Quatrain}} & \multicolumn{1}{c|}{\textbf{Ours}} & \multicolumn{1}{l}{\textbf{\begin{tabular}[c]{@{}l@{}}Improve.\\ CBERT\end{tabular}}} \\ \hline \hline
\textbf{ACS}                                                         & {\color[HTML]{000000} 29:8}                                       & 27.5                                & 40.5                                      & 43.2                                   & 45.9                              & 37.8                                   & \textbf{67.6}                                  & 56.5\%                                                                                                 \\ \hline
\textbf{Arja}                                                        & {\color[HTML]{000000} 8:49}                                       & 80.7                                & 57.9                                      & 87.7                                   & 68.4                              & 63.2                                   & \textbf{94.7}                                  & 8\%                                                                                                    \\ \hline
\textbf{AVATAR}                                                      & {\color[HTML]{000000} 17:37}                                      & 70.4                                & 74.1                                      & 75.9                                   & 64                               & 72.2                                   & \textbf{85.2}                                  & 12.3\%                                                                                                 \\ \hline
\textbf{CapGen}                                                      & {\color[HTML]{000000} 9:41}                                       & 84.0                                & 84.0                                      & 82                                     & 52                                & \textbf{86}                            & 80                                             & -2.4\%                                                                                                 \\ \hline
\textbf{Cardumen}                                                    & {\color[HTML]{000000} 0:9}                                        & 66.7                                & 66.7                                      & 77.8                                   & 37.5                              & 77.8                                   & \textbf{88.9}                                  & 14.3\%                                                                                                 \\ \hline
\textbf{DynaMoth}                                                    & {\color[HTML]{000000} 1:21}                                       & \textbf{95.5}                       & 90.9                                      & \textbf{95.5}                                   & 72.7                              & 68.2                                   & \textbf{95.5}                                  & 0\%                                                                                                    \\ \hline
\textbf{FixMiner}                                                    & {\color[HTML]{000000} 6:19}                                       & 72.0                                & 68.0                                      & 80.0                                   & 64                                & 72                                     & \textbf{84}                                    & 5\%                                                                                                    \\ \hline
\textbf{GenProg}                                                     & {\color[HTML]{000000} 1:24}                                       & 92.0                                & 72.0                                      & \textbf{96.0}                          & 60                                & 68                                     & 92                                             & -4.2\%                                                                                                 \\ \hline
\textbf{HDRepair}                                                    & {\color[HTML]{000000} 4:4}                                        & 62.5                                & 62.5                                      & 75.0                                   & 57.1                             & 62.5                                   & \textbf{87.5}                                  & 16.7\%                                                                                                 \\ \hline
\textbf{Jaid}                                                        & {\color[HTML]{000000} 32:40}                                      & 56.4                                & 61.1                                      & \textbf{69.4}                                   & 48.6                              & 68.1                          & 68.1                                  & -1.9\%                                                                                                 \\ \hline
\textbf{jGenProg}                                                    & {\color[HTML]{000000} 6:33}                                       & 84.6                                & 56.4                                      & 87.2                                   & 69.2                              & 82.1                                   & \textbf{89.7}                                  & 2.9\%                                                                                                  \\ \hline
\textbf{jKali}                                                       & {\color[HTML]{000000} 4:31}                                       & 85.7                                & 65.7                                      & \textbf{94.3}                          & 90.3                             & 74.3                                   & \textbf{94.3}                                  & 0\%                                                                                                    \\ \hline
\textbf{jMutRepair}                                                  & {\color[HTML]{000000} 2:14}                                       & 87.5                                & 68.8                                      & 87.5                                   & 66.7                             & 81.3                                   & \textbf{93.8}                                  & 7.2\%                                                                                                  \\ \hline
\textbf{Kali}                                                        & {\color[HTML]{000000} 2:36}                                       & 89.5                                & 76.3                                      & \textbf{92.1}                          & 84.2                              & 71.1                                   & \textbf{92.1}                                  & 0\%                                                                                                    \\ \hline
\textbf{kPAR}                                                        & {\color[HTML]{000000} 2:32}                                       & 79.4                                & 79.4                                      & \textbf{82.4}                          & 50                                & 64.7                                   & 76.5                                           & -7.2\%                                                                                                 \\ \hline
\textbf{Nopol}                                                       & {\color[HTML]{000000} 6:89}                                       & 92.6                                & 76.8                                      & 69.5                                   & 51.6                              & 71.6                                   & \textbf{93.7}                                  & 34.8\%                                                                                                 \\ \hline
\textbf{RSRepair}                                                    & {\color[HTML]{000000} 2:31}                                       & 84.8                                & 81.8                                      & \textbf{90.9}                          & 75.8                              & 75.8                                   & \textbf{90.9}                                  & 0\%                                                                                                    \\ \hline
\textbf{SequenceR}                                                   & {\color[HTML]{000000} 10:45}                                      & 78.3                       & \textbf{80.0}                                      & 61.8                                   & 74.5                              & 67.3                                   & 72.7                                           & 17.6\%                                                                                                 \\ \hline
\textbf{SimFix}                                                      & {\color[HTML]{000000} 16:42}                                      & 67.2                                & 75.9                                      & 70.7                                   & 60.3                              & 69                                     & \textbf{77.6}                                  & 9.8\%                                                                                                  \\ \hline
\textbf{SketchFix}                                                   & {\color[HTML]{000000} 5:7}                                        & 50.0                                & 66.7                                      & 50.0                                   & 83.3                              & 33.3                                   & \textbf{91.7}                                  & 83.4\%                                                                                                 \\ \hline
\textbf{SOFix}                                                       & {\color[HTML]{000000} 10:1}                                       & 27.3                                & 54.5                                      & 27.3                                   & \textbf{72.7}                    & 45.5                                   & 54.5                                           & 99.6\%                                                                                                 \\ \hline
\textbf{TBar}                                                        & {\color[HTML]{000000} 7:33}                                       & 85.0                                & 75.0                                      & \textbf{87.5}                          & 52.5                              & 75                                     & 85                                             & -2.9\%                                                                                                 \\ \hline \hline
\textbf{Average}                                                     & -                                                                 & 73.6                                & 69.8                                      & 76.5                                   & 63.7                              & 67.6                                   & \textbf{84.4}                                  & 15.9\%                                                                                                 \\ \hline
\textbf{\begin{tabular}[c]{@{}l@{}}Weighted \\ Average\end{tabular}} & -                                                                 & 76.2                                & 70.8                                      & 76.9                                   & 62.3                              & 69.3                                   & \textbf{84.0}                                  & 14.3\%                                                                                                 \\ \hline
\end{tabular}
}
\centering
{\footnotesize CBERT: CodeBERT~\cite{CodeBERT}. Tian's: Tian et al.~\cite{tian2020evaluating}.}
\end{table}
\begin{table}[]
\caption{ F1-scores of \toolname and Baselines.}
\vspace{-0.2cm}
\huge
\label{table:f1}
\resizebox{\columnwidth}{!}{%
\begin{tabular}{l|b|r|r|r|r|r|g|c}
\hline
\textbf{\begin{tabular}[c]{@{}l@{}}Target \\ APR\end{tabular}}       & \textbf{\begin{tabular}[c]{@{}c@{}}correct:\\ wrong\end{tabular}} & \multicolumn{1}{c|}{\textbf{Cache}} & \multicolumn{1}{c|}{\textbf{Tian's}} & \multicolumn{1}{c|}{\textbf{CBERT}} & \multicolumn{1}{c|}{\textbf{ODS}} & \multicolumn{1}{c|}{\textbf{Quatrain}} & \multicolumn{1}{c|}{\textbf{Ours}} & \multicolumn{1}{l}{\textbf{\begin{tabular}[c]{@{}l@{}}Improve.\\ CBERT\end{tabular}}} \\ \hline \hline
\textbf{ACS}                                                         & 29:8                                                              & 38.3                                & 35.3                                 & 40.0                                & 28.6                              & 30.3                                   & \textbf{57.1}                      & 42.8\%                                                                                \\ \hline
\textbf{Arja}                                                        & 8:49                                                              & 89.1                                & 70.7                                 & 93.3                                & 78.6                              & 74.1                                   & \textbf{97.0}                      & 4.0\%                                                                                 \\ \hline
\textbf{AVATAR}                                                      & 17:37                                                             & 81.0                                & 81.6                                 & 82.7                                & 71.0                              & 80.0                                   & \textbf{90.0}                      & 8.8\%                                                                                 \\ \hline
\textbf{CapGen}                                                      & 9:41                                                              & 90.1                                & 90.0                                 & 90.1                                & 62.5                              & \textbf{91.8}                          & 86.8                               & -3.7\%                                                                                \\ \hline
\textbf{Cardumen}                                                    & 0:9                                                               & 87.5                                & 80.0                                 & 87.5                                & 54.5                              & 87.5                                   & \textbf{94.1}                      & 7.5\%                                                                                 \\ \hline
\textbf{DynaMoth}                                                    & 1:21                                                              & \textbf{97.7}                       & 95.2                                 & \textbf{97.7}                       & 83.3                              & 81.1                                   & \textbf{97.7}                      & 0.0\%                                                                                 \\ \hline
\textbf{FixMiner}                                                    & 6:19                                                              & 82.1                                & 77.8                                 & 87.2                                & 71.0                              & 82.1                                   & \textbf{89.5}                      & 2.6\%                                                                                 \\ \hline
\textbf{GenProg}                                                     & 1:24                                                              & 93.6                                & 83.7                                 & \textbf{98.0}                       & 75.0                              & 80.0                                   & {\color[HTML]{333333} 95.8}        & -2.2\%                                                                                \\ \hline
\textbf{HDRepair}                                                    & 4:4                                                               & 72.7                                & 66.7                                 & 80.0                                & 66.7                              & 72.7                                   & \textbf{85.7}                      & 7.1\%                                                                                 \\ \hline
\textbf{Jaid}                                                        & 32:40                                                             & 68.0                                & 69.6                                 & \textbf{76.1}                                & 41.3                              & 74.2                          & 71.6                               & -5.9\%                                                                                \\ \hline
\textbf{jGenProg}                                                    & 6:33                                                              & 91.1                                & 71.2                                 & 93                                  & 78.6                              & 89.6                                   & \textbf{94.3}                      & 1.4\%                                                                                 \\ \hline
\textbf{jKali}                                                       & 4:31                                                              & 92.3                                & 78.6                                 & 96.8                                & 94.7                              & 84.7                                   & \textbf{96.9}                      & 0.1\%                                                                                 \\ \hline
\textbf{jMutRepair}                                                  & 2:14                                                              & 93.3                                & 81.5                                 & 93.3                                & 78.3                              & 88.9                                   & \textbf{96.3}                      & 3.2\%                                                                                 \\ \hline
\textbf{Kali}                                                        & 2:36                                                              & 94.4                                & 86.6                                 & \textbf{95.9}                       & 91.4                              & 82.5                                   & \textbf{95.9}                      & 0.0\%                                                                                 \\ \hline
\textbf{kPAR}                                                        & 2:32                                                              & 88.5                                & 88.1                                 & \textbf{90.3}                       & 65.3                              & 77.8                                   & 86.2                               & -4.5\%                                                                                \\ \hline
\textbf{Nopol}                                                       & 6:89                                                              & 95.0                                & 86.7                                 & 81.5                                & 65.7                              & 83.2                                   & \textbf{96.7}                      & 18.7\%                                                                                \\ \hline
\textbf{RSRepair}                                                    & 2:31                                                              & 90.0                                & 89.7                                 & \textbf{95.2}                       & 85.2                              & 85.7                                   & 95.1                               & -0.1\%                                                                                \\ \hline
\textbf{SequenceR}                                                   & 10:45                                                             & \textbf{85.3}                       & 87.1                                 & 72.0                                & 83.3                              & 79.1                                   & 80.0                               & 11.1\%                                                                                \\ \hline
\textbf{SimFix}                                                      & 16:42                                                             & 79.1                                & 82.5                                 & 80.0                                & 70.9                              & 79.5                                   & \textbf{85.1}                      & 6.4\%                                                                                 \\ \hline
\textbf{SketchFix}                                                   & 5:7                                                               & 62.5                                & 66.7                                 & 50.0                                & 85.7                              & 33.3                                   & \textbf{92.3}                      & 84.6\%                                                                                \\ \hline
\textbf{SOFix}                                                       & 10:1                                                              & 20.0                                & 28.6                                 & 20.0                                & \textbf{40.0}                     & 25.0                                   & 28.6                               & 43.0\%                                                                                \\ \hline
\textbf{TBar}                                                        & 7:33                                                              & 91.2                                & 84.4                                 & \textbf{92.8}                       & 61.2                              & 84.8                                   & 90.6                               & -2.4\%                                                                                \\ \hline \hline
\textbf{Average}                                                     & -                                                                 & 81.0                                & 76.5                                 & 81.5                                & 69.7                              & 74.9                                   & \textbf{86.5}                      & 10.1\%                                                                                \\ \hline
\textbf{\begin{tabular}[c]{@{}l@{}}Weighted \\ Average\end{tabular}} & -                                                                 & 83.5                                & 78.8                                 & 83.2                                & 69.07                             & 77.8                                   & \textbf{87.4}                      & 8.6\%                                                                                 \\ \hline
\end{tabular}
}
\centering
{\footnotesize CBERT: CodeBERT~\cite{CodeBERT}, Tian's: Tian et al.~\cite{tian2020evaluating}}
\end{table}
\begin{table}[]
\caption{ AUC scores of \toolname and Baselines.}
\vspace{-0.2cm}
\huge
\label{table:auc}
\resizebox{\columnwidth}{!}{%
\begin{tabular}{l|b|r|r|r|r|r|g|c}
\hline
\textbf{\begin{tabular}[c]{@{}l@{}}Target \\ APR\end{tabular}}       & \textbf{\begin{tabular}[c]{@{}c@{}}correct:\\ wrong\end{tabular}} & \textbf{Cache}              & \textbf{Tian's}           & \textbf{CBERT}            & \textbf{ODS}              & \textbf{Quatrain}      & \textbf{Ours}                                             & \multicolumn{1}{l}{\textbf{\begin{tabular}[c]{@{}l@{}}Improve.\\ CBERT\end{tabular}}} \\ \hline \hline
\textbf{ACS}                                                         & 29:8                                                              & 74.9                        & 62.9                      & 73.3                      & 40.3                      & 53.9                   & \textbf{76.7}                                             & 4.6\%                                                                                 \\ \hline
\textbf{Arja}                                                        & 8:49                                                              & 79.3                        & 65.6                      & 77.6                      & 76.5                      & 67.1                   & \textbf{89.3}                                             & 15.1\%                                                                                \\ \hline
\textbf{AVATAR}                                                      & 17:37                                                             & 61.2                        & 67.7                      & \textbf{83.6}             & 66.1                      & 72.2                   & 82.7                                                      & -1.1\%                                                                                \\ \hline
\textbf{CapGen}                                                      & 9:41                                                              & 56.6                        & \textbf{93.5}             & 75.9                      & 57.5                      & 84.6                   & 76.4                                                      & 0.7\%                                                                                 \\ \hline
\textbf{Cardumen}                                                    & 0:9                                                               & \multicolumn{1}{r|}{-}      & \multicolumn{1}{r|}{-}    & \multicolumn{1}{r|}{-}    & \multicolumn{1}{r|}{-}    & -                          & -   & -                                                                                   \\ \hline
\textbf{DynaMoth}                                                    & 1:21                                                              & 95.2                        & 0                         & 81                        & \textbf{100}                       & 28.6                   & {\color[HTML]{000000} \textbf{100}}                       & 23.5\%                                                                                \\ \hline
\textbf{FixMiner}                                                    & 6:19                                                              & 52.6                        & 73.7                      & \textbf{87.7}             & 59.6                      & 68.4                   & {\color[HTML]{000000} 77.2}                               & -12.0\%                                                                               \\ \hline
\textbf{GenProg}                                                     & 1:24                                                              & {\color[HTML]{000000} 15.5} & 70.8             & 50                        & 0                         & \textbf{100}                    & {\color[HTML]{000000} 16.7}                               & -66.6\%                                                                               \\ \hline
\textbf{HDRepair}                                                    & 4:4                                                               & 37.5                        & 81.2                      & 56.2                      & 83.3                      & 81.3                   & {\color[HTML]{000000} \textbf{100}}                       & 77.9\%                                                                                \\ \hline
\textbf{Jaid}                                                        & 32:40                                                             & 54.3                        & 63.7                      & 70.9                      & 44.7                      & 70                     & {\color[HTML]{000000} \textbf{72.6}}                      & 2.4\%                                                                                 \\ \hline
\textbf{jGenProg}                                                    & 6:33                                                              & \textbf{88.4}               & 38.4                      & 63.1                      & 85.6                      & 64.6                   & {\color[HTML]{000000} 82.8}                               & 31.2\%                                                                                \\ \hline
\textbf{jKali}                                                       & 4:31                                                              & 83.1                        & 37.1                      & 96.8                      & 49.1                      & 54.8                   & {\color[HTML]{000000} \textbf{99.2}}                      & 2.5\%                                                                                 \\ \hline
\textbf{jMutRepair}                                                  & 2:14                                                              & 50                          & 46.4                      & 82.1                      & 63.5                      & 64.3                   & {\color[HTML]{000000} \textbf{92.9}}                      & 13.2\%                                                                                \\ \hline
\textbf{Kali}                                                        & 2:36                                                              & 43.1                        & 31.9                      & \textbf{88.9}             & 29.2                      & 56.9                   & {\color[HTML]{000000} 77.8}                               & -12.5\%                                                                               \\ \hline
\textbf{kPAR}                                                        & 2:32                                                              & 53.1                        & 54.7                      & \textbf{68.8}             & 64.1                      & 62.5                   & {\color[HTML]{000000} 56.2}                               & -18.3\%                                                                               \\ \hline
\textbf{Nopol}                                                       & 6:89                                                              & 42.7                        & 38.4                      & 60.9                      & \textbf{69.5}             & 46.4                   & {\color[HTML]{000000} 54.1}                               & -11.2\%                                                                               \\ \hline
\textbf{RSRepair}                                                    & 2:31                                                              & 83.9                        & 74.2                      & 61.3                      & 87.1                      & 80.6                   & {\color[HTML]{000000} \textbf{88.7}}                      & 44.7\%                                                                                \\ \hline
\textbf{SequenceR}                                                   & 10:45                                                             & 60.8                        & 74.2                      & 72.7                      & 59.3                      & 57.3                   & {\color[HTML]{000000} \textbf{78.2}}                      & 7.6\%                                                                                 \\ \hline
\textbf{SimFix}                                                      & 16:42                                                             & 50.4                        & 74.1                      & 68.9                      & 65.8                      & 69.2                   & {\color[HTML]{000000} \textbf{79.2}}                      & 14.9\%                                                                                \\ \hline
\textbf{SketchFix}                                                   & 5:7                                                               & 65.7                        & 68.6                      & 60                        & 80                        & 54.3                   & {\color[HTML]{000000} \textbf{94.3}}                      & 57.2\%                                                                                \\ \hline
\textbf{SOFix}                                                       & 10:1                                                              & 70                          & 80                        & 90                        & 95                        & 80                     & {\color[HTML]{000000} \textbf{100}}                       & 11.1\%                                                                                \\ \hline
\textbf{TBar}                                                        & 7:33                                                              & 72.7                        & 70.1                      & 63.2                      & 80.3                      & 65.8                   & {\color[HTML]{000000} \textbf{92.4}}                      & 46.2\%                                                                                \\ \hline \hline
\textbf{Average}                                                     & -                                                                 & \multicolumn{1}{r|}{61.5}   & \multicolumn{1}{r|}{60.3} & \multicolumn{1}{r|}{73.0} & \multicolumn{1}{r|}{64.6} & 65.8                   & \textbf{80.4} & 11.0\%                                                                                \\ \hline
\textbf{\begin{tabular}[c]{@{}l@{}}Weighted \\ Average\end{tabular}} & -                                                                 & \multicolumn{1}{r|}{60.8}   & \multicolumn{1}{r|}{59.9} & \multicolumn{1}{r|}{72.5} & \multicolumn{1}{r|}{62.5} & 64.5                   & \textbf{76.9} & 16.5\%                                                                                \\ \hline
\end{tabular}
}
\centering
{\footnotesize CBERT: CodeBERT~\cite{CodeBERT}, Tian's: Tian et al.~\cite{tian2020evaluating}}
\end{table}

\begin{table}[b]
\vspace{-0.4cm}
\caption{Comparison of \toolname with ODS on APR data where ODS inference fails in some patches.}
\vspace{-0.2cm}
\label{table:compare_ods}
\centering
\resizebox{0.95\columnwidth}{!}{%
\begin{tabular}{l|c|ccc|ccg}
\hline
\multirow{2}{*}{\textbf{\begin{tabular}[c]{@{}l@{}}Target APR\\ Tools\end{tabular}}} & \multirow{2}{*}{\textbf{cor:incor}} & \multicolumn{3}{c|}{\textbf{ODS}} & \multicolumn{3}{c}{\textbf{LLM4PatchCorrect}} \\ \cline{3-8} 
 &  & \multicolumn{1}{c|}{
 \textbf{Acc}} & \multicolumn{1}{c|}{\textbf{F1}} & \textbf{AUC} & \multicolumn{1}{c|}{\textbf{Acc}} & \multicolumn{1}{c|}{\textbf{F1}} &  \multicolumn{1}{c}{\textbf{AUC}} \\ \hline
\textbf{AVATAR} & 16:34 & \multicolumn{1}{c|}{64} & \multicolumn{1}{c|}{71} & 66.1 & \multicolumn{1}{g|}{\textbf{86.0}} & \multicolumn{1}{g|}{\textbf{90.4}} &\textbf{84.7}  \\ \hline
\textbf{Cardumen} & 0:8 & \multicolumn{1}{c|}{37.5} & \multicolumn{1}{c|}{54.5} & - & \multicolumn{1}{g|}{\textbf{87.5}} & \multicolumn{1}{g|}{\textbf{93.3}} &-  \\ \hline
\textbf{HDRepair} & 4:3 & \multicolumn{1}{r|}{57.1} & \multicolumn{1}{l|}{66.7} & \multicolumn{1}{c|}{83.3} & \multicolumn{1}{g|}{\textbf{85.7}} & \multicolumn{1}{g|}{\textbf{80.0}} &\multicolumn{1}{g}{\textbf{100}} \\ \hline
\textbf{jKali} & 4:27 & \multicolumn{1}{c|}{90.3} & \multicolumn{1}{c|}{94.7} & 49.1 & \multicolumn{1}{g|}{\textbf{93.5}} & \multicolumn{1}{g|}{\textbf{96.4}} & \textbf{99.1} \\ \hline
\textbf{jMutRepair} & 2:13 & \multicolumn{1}{c|}{66.7} & \multicolumn{1}{c|}{78.3} & 63.5 & \multicolumn{1}{g|}{\textbf{93.3}} & \multicolumn{1}{g|}{\textbf{96.0}} &\textbf{92.3}  \\ \hline
\end{tabular}
}
\end{table}

\revised{Moreover, we observed that the accuracy and F1 scores for SOFix are relatively low compared to other APR tools. This is primarily due to a significant imbalance in the test set, where there are 10 correct patches for every 1 overfitting patch, resulting in a labeling ratio of 1:10. This highly imbalanced test set presents a challenge not only for our approach but also for all methods.}

\revised{
Additionally, it's worth noting that during the execution of ODS, we encountered challenges in performing inference for certain instances when the target APR tools are jKali (4 failing test patches), AVATAR (4 failing test patches), jMutRepair (1 failing test patch), HDRepair (1 failing test patch), and Cardumen (1 failing test patch). Please refer to our replication package for details on all the failing test patches.
Consequently, the performance metrics of ODS presented in Tables~\ref{table:acc}--\ref{table:auc} are calculated based on the successfully tested patches only. To ensure a fair comparison, we also evaluate \toolname using the identical test sets that ODS can successfully infer.
As there are differences when the target APR tools are jKali, AVATAR, jMutRepair, HDRepair, and Cardumen, we only re-evaluate \toolname on those APR tools.
Table~\ref{table:compare_ods} displays the performance of ODS and \toolname on identical test sets. Our approach consistently and significantly outperforms ODS.
}

\vspace{0.1cm}
\noindent\textbf{Comparison with Patch-Sim.}
\revised{We also conducted a comparison between our \toolname and the dynamic-based APCA approach Patch-Sim~\cite{xiong2018identifying}.
Because implementing Patch-Sim on new datasets is complex, we relied on prediction results from an empirical study by Wang et al.~\cite{Wang2020AutomatedPC} for comparison. Specifically, we used the dataset from their study to evaluate \toolname in the cross-tool validation setting and compared Patch-Sim's performance in the same dataset under the same cross-tool validation.
}

\revised{As shown in Table~\ref{table:compare_patchsim}, our approach significantly outperforms Patch-Sim on average of all the target APR tools. Specifically, \toolname showed 68.1\% ($\frac{82.2-48.9}{48.9}$), 64.6\% ($\frac{84.3-51.2}{51.2}$), and 33.5\% ($\frac{80.1-60.0}{60.0}$) improvements against Patch-Sim on average in terms of Accuracy, F1-score, and AUC.}

\begin{table}[t]
\caption{Comparison of \toolname with Patch-Sim on Dataset from~\cite{Wang2020AutomatedPC}.}
\vspace{-0.2cm}
\label{table:compare_patchsim}
\centering
\resizebox{0.85\columnwidth}{!}{%
\begin{tabular}{l|ccc|ccg}
\hline
\multirow{2}{*}{\textbf{\begin{tabular}[c]{@{}l@{}}Target APR\\ Tools\end{tabular}}} & \multicolumn{3}{c|}{\textbf{Patch-Sim}} & \multicolumn{3}{c}{\textbf{LLM4PatchCorrect}} \\ \cline{2-7} 
 & \multicolumn{1}{c|}{\textbf{Acc}} & \multicolumn{1}{c|}{\textbf{F1}} & \multicolumn{1}{c|}{\textbf{AUC}} & \multicolumn{1}{g|}{\textbf{Acc}} & \multicolumn{1}{g|}{\textbf{F1}} & \multicolumn{1}{g}{\textbf{AUC}} \\ \hline
\textbf{ACS} & \multicolumn{1}{r|}{\textbf{65.0}} & \multicolumn{1}{r|}{22.2} & \textbf{53.3} & \multicolumn{1}{g|}{45.0} & \multicolumn{1}{g|}{\textbf{52.2}} & 45.1 \\ \hline
\textbf{Arja} & \multicolumn{1}{r|}{35.1} & \multicolumn{1}{r|}{46.4} & 55.4 & \multicolumn{1}{g|}{\textbf{94.7}} & \multicolumn{1}{g|}{\textbf{97.1}} & \textbf{81.2} \\ \hline
\textbf{AVATAR} & \multicolumn{1}{r|}{50.0} & \multicolumn{1}{r|}{54.2} & 54.0 & \multicolumn{1}{g|}{\textbf{77.8}} & \multicolumn{1}{g|}{\textbf{85.4}} & \textbf{77.7} \\ \hline
\textbf{CapGen} & \multicolumn{1}{r|}{61.2} & \multicolumn{1}{r|}{67.5} & 60.1 & \multicolumn{1}{g|}{\textbf{70.1}} & \multicolumn{1}{g|}{\textbf{73.7}} & \textbf{81.4} \\ \hline
\textbf{Cardumen} & \multicolumn{1}{r|}{41.7} & \multicolumn{1}{r|}{53.3} & - & \multicolumn{1}{g|}{\textbf{91.7}} & \multicolumn{1}{g|}{\textbf{94.7}} & \textbf{-} \\ \hline
\textbf{DynaMoth} & \multicolumn{1}{r|}{31.8} & \multicolumn{1}{r|}{44.4} & 64.3 & \multicolumn{1}{g|}{\textbf{95.5}} & \multicolumn{1}{g|}{\textbf{97.7}} & \textbf{95.2} \\ \hline
\textbf{FixMiner} & \multicolumn{1}{r|}{50.0} & \multicolumn{1}{r|}{46.7} & 55.0 & \multicolumn{1}{g|}{\textbf{78.1}} & \multicolumn{1}{g|}{\textbf{83.7}} & \textbf{82.5} \\ \hline
\textbf{GenProg} & \multicolumn{1}{r|}{35.7} & \multicolumn{1}{r|}{50.0} & 66.7 & \multicolumn{1}{g|}{\textbf{96.4}} & \multicolumn{1}{g|}{\textbf{98.1}} & \textbf{100} \\ \hline
\textbf{Jaid} & \multicolumn{1}{r|}{\textbf{63.0}} & \multicolumn{1}{r|}{59.5} & \textbf{62.9} & \multicolumn{1}{g|}{56.8} & \multicolumn{1}{g|}{\textbf{61.5}} & 60.9 \\ \hline
\textbf{jGenProg} & \multicolumn{1}{r|}{47.4} & \multicolumn{1}{r|}{61.5} & 41.7 & \multicolumn{1}{g|}{\textbf{84.2}} & \multicolumn{1}{g|}{\textbf{90.9}} & \textbf{66.7} \\ \hline
\textbf{jKali} & \multicolumn{1}{r|}{45.8} & \multicolumn{1}{r|}{58.1} & 70.5 & \multicolumn{1}{g|}{\textbf{95.8}} & \multicolumn{1}{g|}{\textbf{97.7}} & \textbf{97.7} \\ \hline
\textbf{jMutRepair} & \multicolumn{1}{r|}{59.1} & \multicolumn{1}{r|}{69.0} & 59.4 & \multicolumn{1}{g|}{\textbf{90.9}} & \multicolumn{1}{g|}{\textbf{94.1}} & \textbf{95.3} \\ \hline
\textbf{Kali} & \multicolumn{1}{r|}{31.7} & \multicolumn{1}{r|}{43.8} & 64.0 & \multicolumn{1}{g|}{\textbf{95.0}} & \multicolumn{1}{g|}{\textbf{97.4}} & \textbf{78.9} \\ \hline
\textbf{kPAR} & \multicolumn{1}{r|}{48.3} & \multicolumn{1}{r|}{59.7} & 53.0 & \multicolumn{1}{g|}{\textbf{86.7}} & \multicolumn{1}{g|}{\textbf{92.2}} & \textbf{79.8} \\ \hline
\textbf{Nopol} & \multicolumn{1}{r|}{33.3} & \multicolumn{1}{r|}{47.4} & \textbf{65.5} & \multicolumn{1}{g|}{\textbf{96.7}} & \multicolumn{1}{g|}{\textbf{98.3}} & 51.7 \\ \hline
\textbf{RSRepair} & \multicolumn{1}{r|}{30.0} & \multicolumn{1}{r|}{41.7} & 63.2 & \multicolumn{1}{g|}{\textbf{95.0}} & \multicolumn{1}{g|}{\textbf{97.4}} & \textbf{92.1} \\ \hline
\textbf{SequenceR} & \multicolumn{1}{r|}{49.3} & \multicolumn{1}{r|}{59.1} & 51.8 & \multicolumn{1}{g|}{\textbf{70.4}} & \multicolumn{1}{g|}{\textbf{76.9}} & \textbf{82.2} \\ \hline
\textbf{SimFix} & \multicolumn{1}{r|}{50.0} & \multicolumn{1}{r|}{50.7} & 57.0 & \multicolumn{1}{g|}{\textbf{77.3}} & \multicolumn{1}{g|}{\textbf{84.8}} & \textbf{77.1} \\ \hline
\textbf{SketchFix} & \multicolumn{1}{r|}{\textbf{70.8}} & \multicolumn{1}{r|}{36.4} & 58.4 & \multicolumn{1}{g|}{\textbf{70.8}} & \multicolumn{1}{g|}{\textbf{58.8}} & \textbf{75.6} \\ \hline
\textbf{SOFix} & \multicolumn{1}{r|}{\textbf{73.9}} & \multicolumn{1}{r|}{\textbf{50.0}} & 85.0 & \multicolumn{1}{g|}{\textbf{73.9}} & \multicolumn{1}{g|}{\textbf{50.0}} & \textbf{95.0} \\ \hline
\textbf{TBar} & \multicolumn{1}{r|}{52.9} & \multicolumn{1}{r|}{53.5} & 59.3 & \multicolumn{1}{g|}{\textbf{82.9}} & \multicolumn{1}{g|}{\textbf{88.0}} & \textbf{86.2} \\ \hline
\textbf{Average} & \multicolumn{1}{c|}{48.9} & \multicolumn{1}{c|}{51.2} & \multicolumn{1}{c|}{60.0} & \multicolumn{1}{g|}{\textbf{82.2}} & \multicolumn{1}{g|}{\textbf{84.3}} & \multicolumn{1}{g}{\textbf{80.1}} \\ \hline
\textbf{\begin{tabular}[c]{@{}l@{}}Weighted \\ Average\end{tabular}} & \multicolumn{1}{c|}{49.0} & \multicolumn{1}{c|}{52.8} & \multicolumn{1}{c|}{59.3} & \multicolumn{1}{g|}{\textbf{80.5}} & \multicolumn{1}{g|}{\textbf{83.9}} & \multicolumn{1}{g}{\textbf{79.4}} \\ \hline
\end{tabular}
}
\end{table}

\begin{figure}[b]
\centering
\vspace{-0.4cm}
\subfigure[CodeBERT]{
\begin{minipage}[t]{0.4\linewidth}
\centering
\includegraphics[width=\textwidth]{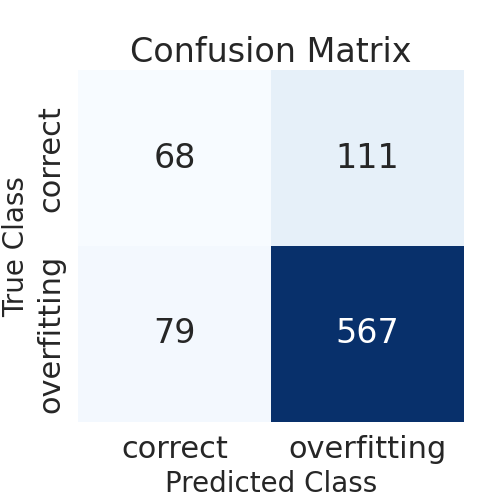}
\end{minipage}%
}%
\subfigure[\toolname]{
\begin{minipage}[t]{0.4\linewidth}
\centering
\includegraphics[width=\textwidth]{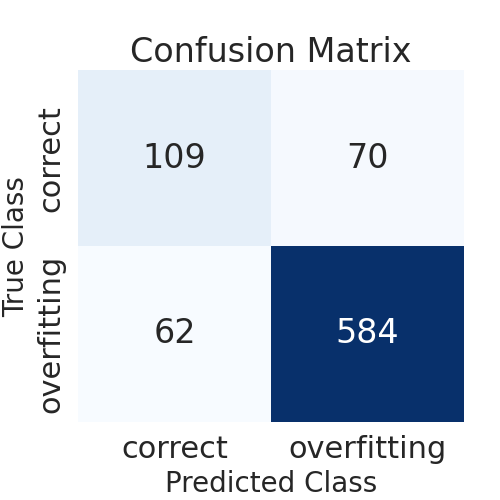}
\end{minipage}%
}%
\centering
\captionsetup{justification=raggedright}
\vspace{-0.2cm}
\caption{Confusion Matrices of \toolname and the best-performing baseline CodeBERT.}
\label{confusion}
\end{figure}

\vspace{0.1cm}
\noindent\textbf{Confusion Matrix Analysis.}
\revised{
Figure~\ref{confusion} compares the confusion matrices of \toolname and the best-performing baseline (i.e. CodeBERT). For ease of analysis, we put the predictions of each test set in the ``cross-tool'' setting together and compute the confusion matrices. In total, we have 825 patches generated by 22 APR tools. Please note that though the whole dataset has 1,179 patches, only 825 of them are generated by APR tools and the other patches are contributed by developers. Among the 825 generated patches, 646 patches are overfitting and only 179 patches are correct.
The correct patch ratio of APR-generated patches is only 21.7\% ($\frac{179}{179+646}$) without using APCA techniques.
Using \toolname, it can filter out 584 overfitting patches (i.e., 90.4\% of overfitting patches) as shown in Figure~\ref{confusion} (b).
The remaining patches are predicted as correct by \toolname and the correct patch ratio of the remaining patches is increased from 21.7\% to 63.7\% ($\frac{109}{109+62}$).
For CodeBERT, as shown in Figure~\ref{confusion}a, it can filter out 567 overfitting patches (i.e., 87.7\% of overfitting patches) but the correct patch ratio of the remaining patches is only 46.3\% ($\frac{68}{68+79}$). 
\revised{\toolname has a 37.6\% ($\frac{63.7\%-46.3\%}{46.3\%}$) relative improvement over CodeBERT in terms of the correct patch ratio of the remaining patches.}
Though APCA techniques can help to reduce the number of overfitting patches, they inevitably delete some correct patches at the same time. Figure~\ref{confusion} (a) also shows that CodeBERT wrongly predicts 111 correct patches as ``overfitting'' while the corresponding number for \toolname is \revised{70}. It indicates that \toolname makes fewer mistakes than CodeBERT. 
}

\begin{tcolorbox}[enhanced,width=\linewidth,drop fuzzy shadow southwest,colback=gray!10]
    \textbf{Answer to RQ1}:  
    \revised{\toolname significantly outperforms the baseline techniques. By utilizing both the labeled patches from existing APR tools and additional guiding information from the bug benchmark, our approach accurately predicts the correctness of patches generated by the target APR tool, even in the absence of labeled patches specific to that tool.}
\end{tcolorbox}

\subsection{RQ2. Ablation Study}
\label{subsection:rq2}

\noindent
\textbf{\revised{Method.}}
\revised{
To delineate the contributions of each component of \toolname, we conduct an ablation study. The ablation studies are based on three groups of guiding information:
\begin{itemize}
    \item Bug Information: Bug descriptions and execution traces;
    \item Test Information: Failing test cases and test coverages;
    \item Retrieved Patches: Patches retrieved from the existing APR tools (the training set).
\end{itemize}
}

\revised{For the ablation study, we initially employed only the LLM (Starcoder-7B) without any guiding information. Subsequently, we combined each category of the three groups of guiding information with the LLM (Starcoder-7B) to perform the APCA task. Finally, we incorporate all the guiding information (the full model) to complete the task.}

\begin{table}[t]
\caption{Results of the ablation study in terms of Accuracy, F1-score, and AUC, on average of all APR tools.}
\label{table:ablation}
\huge
\centering
\resizebox{\columnwidth}{!}{%
\begin{tabular}{l|c|c|c|c|c|c|c|c|c}
\hline
\textbf{Ablation}                                                                & \textbf{LLM} & \textbf{\begin{tabular}[c]{@{}c@{}}bug \\ descri.\end{tabular}} & \textbf{\begin{tabular}[c]{@{}c@{}}exe.\\ traces\end{tabular}} & \textbf{\begin{tabular}[c]{@{}c@{}}test\\ cases\end{tabular}} & \textbf{\begin{tabular}[c]{@{}c@{}}test\\ cover.\end{tabular}} & \textbf{\begin{tabular}[c]{@{}c@{}}retrieved\\ patches\end{tabular}} & \textbf{Acc.} & \textbf{F1} & \textbf{AUC} \\ \hline
\textbf{No Design}                                                               & \textbf{\begin{tikzpicture}
    \fill[green] (0,0) circle (0.3cm);
\end{tikzpicture}}   &                                                                 &                                                                &                                                               &                                                                  &                                                                      & 75.4          & 83.1        & 41.8         \\ \hline
\multirow{2}{*}{\textbf{\begin{tabular}[c]{@{}l@{}}Info.of\\ Bug/Tests\end{tabular}}} & \textbf{\begin{tikzpicture}
    \fill[green] (0,0) circle (0.3cm);
\end{tikzpicture}}   & \textbf{\begin{tikzpicture}
    \fill[green] (0,0) circle (0.3cm);
\end{tikzpicture}}                                                      & \textbf{\begin{tikzpicture}
    \fill[green] (0,0) circle (0.3cm);
\end{tikzpicture}}                                                     & \textbf{}                                                     & \textbf{}                                                        & \textbf{}                                                            & 75.7          & 83.5        & 45.8         \\ \cline{2-10} 
                                                                                 & \textbf{\begin{tikzpicture}
    \fill[green] (0,0) circle (0.3cm);
\end{tikzpicture}}   & \textbf{}                                                       & \textbf{}                                                      & \textbf{\begin{tikzpicture}
    \fill[green] (0,0) circle (0.3cm);
\end{tikzpicture}}                                                    & \textbf{\begin{tikzpicture}
    \fill[green] (0,0) circle (0.3cm);
\end{tikzpicture}}                                                       & \textbf{}                                                            & 76.0          & 83.7        & 43.3         \\ \hline
\textbf{\begin{tabular}[c]{@{}l@{}}Labeled \\ Patches\end{tabular}}              & \textbf{\begin{tikzpicture}
    \fill[green] (0,0) circle (0.3cm);
\end{tikzpicture}}   & \textbf{}                                                       & \textbf{}                                                      & \textbf{}                                                     & \textbf{}                                                        & \textbf{\begin{tikzpicture}
    \fill[green] (0,0) circle (0.3cm);
\end{tikzpicture}}                                                           & \underline{84.0}          & \underline{86.3}        & \underline{77.2}         \\ \hline
\rowcolor{green!10} \textbf{Full}                                                                    & \textbf{\begin{tikzpicture}
    \fill[green] (0,0) circle (0.3cm);
\end{tikzpicture}}   & \textbf{\begin{tikzpicture}
    \fill[green] (0,0) circle (0.3cm);
\end{tikzpicture}}                                                      & \textbf{\begin{tikzpicture}
    \fill[green] (0,0) circle (0.3cm);
\end{tikzpicture}}                                                     & \textbf{\begin{tikzpicture}
    \fill[green] (0,0) circle (0.3cm);
\end{tikzpicture}}                                                    & \textbf{\begin{tikzpicture}
    \fill[green] (0,0) circle (0.3cm);
\end{tikzpicture}}                                                       & \textbf{\begin{tikzpicture}
    \fill[green] (0,0) circle (0.3cm);
\end{tikzpicture}}                                                           & \textbf{84.4}          & \textbf{86.5}        & \textbf{80.4}         \\ \hline
\end{tabular}
\vspace{-0.4cm}
}
\end{table}

\vspace{0.2cm}
\noindent
\textbf{Results.}
\revised{
We initially investigated the effects of integrating guiding information from the bug benchmark. As depicted in Table~\ref{table:ablation}, on average, utilizing bug-related information, including bug descriptions and execution traces, consistently yields improvements across all metrics. These enhancements demonstrate a relative improvement of up to 9.6\% compared to solely relying on the large language model (e.g., Starcoder-7B). Additionally, information related to test cases (i.e., failing test cases and test coverages) also consistently leads to improvements across all metrics compared to relying solely on the large language model.
}

\revised{Additionally, we examined the contribution of the contrastive learning-based patch retrieval module, which retrieves patches from the training set. As shown in Table~\ref{table:ablation}, on average, the patch retrieval module leads to relative improvements of 11.4\%, 3.9\%, and 84.7\% in Accuracy, F1-score, and AUC, respectively, compared to relying solely on the large language model.
Moreover, we discovered that the guiding information from the bug benchmark and the retrieved patches complement each other, resulting in improved results for the full model of \toolname.}

\begin{tcolorbox}[enhanced,width=\linewidth,drop fuzzy shadow southwest,colback=gray!10]
\textbf{Answer to RQ2}: 
\revised{All modules contribute to the effectiveness of \toolname. Incorporating all designs, our approach achieves relative improvements of 11.9\%, 4.1\%, and 92.3\% in Accuracy, F1-score, and AUC, respectively, compared to relying solely on the large language model.}
\end{tcolorbox}

\begin{figure*}[t] 
\centering 
\includegraphics[width=0.75\linewidth]{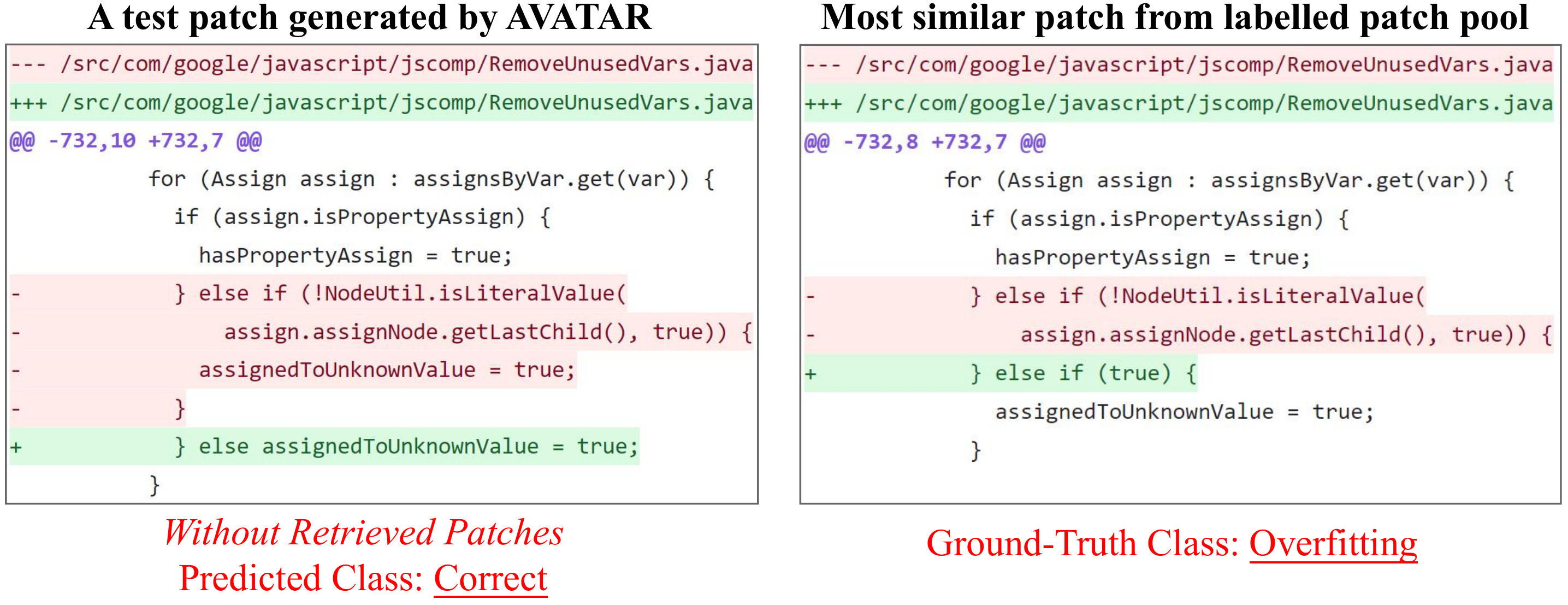} 
\vspace{-0.2cm}
\caption{A test patch generated by AVATAR and the predictions given by \toolname with and without the instance-wise demonstration.}
\label{fig:example} 
\vspace{-0.2cm}
\end{figure*}

\section{Discussion}
\label{sec:discussion}

 \subsection{Case Study}

\revised{The ablation study (RQ2) revealed that the contrastive learning-based patch retrieval module, utilized to acquire semantically similar patches from the training set, plays the most significant role in ensuring the accurate prediction of \toolname.
In this discussion, we give a case study to demonstrate how our proposed contrastive learning-based patch retrieval module would enhance the \toolname.
To achieve this, we compare the prediction made by the complete \toolname model with that of its variant that excludes the retrieved labeled patches when evaluating the correctness of a given test patch.}

\revised{Figure~\ref{fig:example} illustrates an example of a test patch extracted from the {\tt closure-compiler} project, aiming to fix the Closure-45 bug in Defects4J, along with the most similar patch retrieved from the training set.}
Such a test patch (the left-hand side code) tries to repair the program by removing the condition of the {\tt else if} statement.
\revised{\toolname without retrieved labeled patches predicts the input test patch as ``correct''. But this test patch, in fact, is an overfitting patch.}
\revised{However, using the full \toolname (including the retrieved labeled patches) the prediction is changed to ``overfitting'' which is the ground truth.}
The most similar patch retrieved from the training set (the right-hand code) also tries to repair the program by relaxing the condition of the {\tt else if} statement but this patch is already identified as an overfitting patch. 
As these two patches are functionally equivalent, the label of the most similar patch is valuable for the LLM to judge the test patch.

In general, we observe that the contrastive learning-based patch retrieval module could provide relevant information to the LLM and correct the potential errors in its initial predictions. 

\revised{Moreover, please kindly note that the patch on the right in Figure~\ref{fig:example} is retrieved from the training set by our approach, but it is also utilized in the training of all baselines like Cache. In other words, this patch on the right is ``seen" by all baselines during their model training, ensuring fairness in the evaluation of baselines.
}

\subsection{Effectiveness on the Bears Benchmark}\label{sec:discussion_bear}

\begin{table}[t]
\caption{ Accuracy of \toolname and Baselines on the Bears benchmark~\cite{bears}.}
\label{table:bear_acc}
\centering
\huge
\resizebox{0.9\columnwidth}{!}{%
\begin{tabular}{lcccccg}
\hline
\textbf{Accuracy} & \textbf{Tain et al.} & \textbf{CodeBERT} & \textbf{ODS} & \textbf{Quatrain} & \textbf{Cache} & \textbf{Ours} \\ \hline
Arja              & 9.0                & 4.3               & 3.7            & 7.3    & 3.7            & \textbf{85.0}                        \\
GenProg           & 73.4             & 88.3              & 47.8            & 99.5      & \textbf{100}         & \textbf{100}                       \\
Kali              & 16.7             & 16.7              & 0.0            & \textbf{83.3}    & 40.0           & \textbf{83.3}                     \\
RSRepair          & \textbf{100}             & \textbf{100}               & \textbf{100}            & \textbf{100}    & \textbf{100}            & \textbf{100}                       \\ \hline
\textbf{Average}      & 49.8             & 52.3              & 37.9            & 72.5       & 60.9       & \textbf{92.1}                      \\ \hline
\textbf{\begin{tabular}[c]{@{}l@{}}Weighted \\ Average\end{tabular}}          & 36.4             & 39.9              & 22.5            & 47.1      & 44.7        & \textbf{91.3}                      \\ \hline
\end{tabular}
}
\end{table}

\begin{table}[t]
\caption{ F1-score of \toolname and Baselines on the Bears benchmark~\cite{bears}.}
\label{table:bear_f1}
\centering
\huge
\resizebox{0.9\columnwidth}{!}{%
\begin{tabular}{lcccccg}
\hline
\textbf{F1} & \textbf{Tain et al.} & \textbf{CodeBERT} & \textbf{ODS} & \textbf{Quatrain} & \textbf{Cache} & \textbf{Ours} \\ \hline
Arja        & 16.5             & 8.3               & 7.1            & 13.6   & 7.1           & \textbf{91.8}                      \\
GenProg     & 84.7             & 93.8              & 64.6            & 99.8   & \textbf{100}          & \textbf{100}                       \\
Kali        & 28.6             & 28.6              & 0            & 90.3   & 57.1           & \textbf{90.9}                      \\
RSRepair    & \textbf{100}              & \textbf{100}                & \textbf{100}            & \textbf{100}      & \textbf{100}          & \textbf{100}                       \\ \hline
\textbf{Average}      & 57.5             & 57.7              & 42.9            & 76.1       & 66.1       & \textbf{95.7}                      \\ \hline
\textbf{\begin{tabular}[c]{@{}l@{}}Weighted \\ Average\end{tabular}}          & 45.5             & 44.6              & 31.4            & 50.8      & 46.9        & \textbf{95.2}                      \\ \hline
\end{tabular}
}
\end{table}

\revised{The main experiments are conducted based on the patches for Defects4J benchmark~\cite{defect4j}. To ensure that our approach can work beyond the Defects4J
benchmark, we did additional experiments on the Bears benchmark~\cite{bears}. }

\revised{
The patches used in this discussion are sourced from~\cite{Ye2022AutomatedCO} and there are patches generated by four APR tools: Arja, GenProg, Kali, and RSRepair. 
To conduct the experiments, we still follow our cross-tool validation setting, where we consider the patches generated by one APR as the test set and the rest as the training set.
}

\revised{
Table~\ref{table:bear_acc} and Table~\ref{table:bear_f1} demonstrate the effectiveness of both the baselines and our approach in the Bears benchmark. 
The experimental results highlight the exceptional performance of \toolname in the Bears benchmark. On average, \toolname outperformed the best-performing baseline (Quatrain~\cite{quatrain}) by 27.0\% ($\frac{92.1-72.5}{72.5}$) in Accuracy and 25.8\% ($\frac{95.7-76.1}{76.1}$) in F1-score on average.
This underscores the effectiveness of LLM4PatchCorrect in a new dataset beyond the Defects4J.
}

\subsection{\revised{Impact of the LLM Choices}}

\revised{
In this subsection, we aim to discuss the impacts of selecting different LLM models as the backbone model. Prior research~\cite{bert,gpt3,wei2022emergent} suggests that increasing the size of LLMs, such as expanding the pre-training corpus and model parameters, generally leads to improved performance. 
Due to computational resource constraints in academic settings, we limit our selection to LLMs up to 7B parameters. We choose the LLMs up to 7B and we argue that the 7B parameters are sufficiently large for research purposes.
}

\revised{
In addition to the Starcoder-7B model, we also implemented a version of \toolname based on another popular 7B model, named Code LLamma-7B~\cite{codellama}. Furthermore, to validate the relationship between model sizes and effectiveness within the framework of \toolname, we included smaller LLMs such as CodeGen2-3.7B~\cite{nijkamp2023codegen2}, Starcoder-3B~\cite{li2023starcoder}, Starcoder-1B~\cite{li2023starcoder}, and BLOOM-1.7B~\cite{bloom}. The results are presented in Table~\ref{table:impact_llms}, with the highlighted row indicating the original version of \toolname.
}

\revised{
As shown in Table~\ref{table:impact_llms}, we observe that larger model sizes generally exhibit better performance. Nevertheless, we note that Starcoder-1B also exhibits competitive performance despite its relatively small model size.
Moreover, we find that the effectiveness of \toolname remains stable when replacing Starcoder-7B with another popular 7B model; the difference in effectiveness is less than 1\%, indicating that \toolname maintains satisfactory performance with a different LLM.
}

\begin{table}[t]
\centering
\caption{Impacts of different LLMs on \toolname.}
\label{table:impact_llms}
\resizebox{0.9\columnwidth}{!}{%
\begin{tabular}{ll|ccc}
\hline
\textbf{\begin{tabular}[c]{@{}l@{}}Model Size\\ (M)\end{tabular}} & \multicolumn{1}{l|}{\textbf{LLMs}} & \textbf{Acc.}  & \textbf{F1}   & \textbf{AUC}  \\ \hline
\rowcolor{green!10} \textbf{M=7B}                 
  &  Starcoder-7B~\cite{li2023starcoder} (Ours)                       & \textbf{84.4} & \textbf{86.5} & \textbf{80.4} \\
\textbf{}                                                            & CodeLLama-7B~\cite{codellama}                       & 83.7          & 85.6          & \textbf{80.4} \\ \hline
\multirow{2}{*}{\textbf{2B\textless{}M\textless{}7B}}                & CodeGen2-3.7B~\cite{nijkamp2023codegen2}                       & 81.3          & 84.1          & 77.6          \\
                                                                     & Starcoder-3B~\cite{li2023starcoder}                       & 80.1          & 82.6          & 76.6          \\ \hline
\multirow{2}{*}{\textbf{1B\textless{}M\textless{}2B}}                & BLOOM-1.7B~\cite{bloom}                         & 80.4          & 83.2          & 76.4          \\
                                                                     & Starcoder-1B~\cite{li2023starcoder}                       & 81.9          & 85.9          & 73.8          \\ \hline
\end{tabular}
}
\end{table}

\subsection{\revised{Application Scenario}}
\revised{
Our approach is designed to predict the correctness of patches generated by APR tools. Therefore, the application scenario is as follows: it can be used to filter out less trustworthy generated candidates. Before delving into the application scenario of our approach, let us briefly explain the typical usage of APR tools:
\begin{itemize}
    \item Step 1: Given a bug, the APR tool typically generates N candidate patches (e.g., N=100).
    \item Step 2: These candidate patches are applied to the software to determine if they can successfully pass all available tests. If they do, the candidate patch is considered ``plausible".
    \item Step 3: Developers manually inspect all ``plausible" patches to verify whether they are indeed correct patches or just overfitting patches. 
\end{itemize}
Manually verifying the correctness of patches (Step 3) is a time-consuming and tedious task. Our approach aims to assist developers in filtering out plausible patches that are less likely to be correct, i.e., those predicted as ``overfitting" by our approach. This helps reduce the number of plausible patches developers need to manually check, improving efficiency.
}

\subsection{Threats to Validity}

\noindent\textbf{Threats to External Validity.}  
Large pre-trained models diverge depending on different aspects, such as the characteristics of pre-training tasks and the size of pre-training datasets.
As a threat to validity, our study may have a selection bias by considering only several large pre-trained models. 
To mitigate this threat, we conducted preliminary experiments with existing open-source models and kept tracking the models that are not publicly shared online.
Another threat to validity can be dataset selection, as it may deliver bias in the experimental results. The selection, however, is to compare against the state-of-the-art following their settings. 
Researchers~\cite{tian2020evaluating,Wang2020AutomatedPC} reported that the dataset had been checked for its correctness which automatically minimized this threat to validity. 
Finally, the evaluation metrics we borrow sometimes may cause bias depending upon the characteristics of the tasks.
We believe this threat is mitigated as we double-checked, and they are well-known for classification tasks.
Furthermore, We publicly share our implementation and dataset for future comparisons by the research community.

\noindent\textbf{Threats to Internal Validity.}  
The main threat to internal validity lies in the manually crafted prompt that we designed for our model.
We cannot ensure that our prompt is optimal as well as it is impossible to traverse all the potential prompts. 
We mitigate this by following the most common prompts~\cite{Sanh2022MultitaskPT} and we share the prompt in the artifacts for the community to review.

\noindent\textbf{Threats to Construct Validity.}
The large pre-trained model we employ in our study is not perfect, and it may have been under-trained, which can affect its complete effectiveness as we pre-envisioned in the previous Section.
This may imply that our design can boost the tool's effectiveness by capturing more practical features for the assessment. 
We believe our future study can shed light on this threat by considering larger models.

\section{Related Work}
\label{sec:related}

\revised{Many automated patch correctness assessment (APCA) approaches~\cite{Tan2016AntipatternsIS, xiong2018identifying, Yang2017BetterTC, tian2020evaluating,  Ye2022AutomatedCO, Ye2021AutomatedPA, Wang2020AutomatedPC, Xin2017IdentifyingTP, Wen2018ContextAwarePG, Xin2017LeveragingSC, Ye2022AutomatedCO, Le2017S3SA, cache, quatrain, le2023invalidator} have been proposed to conduct the patch correctness assessment.
The APCA approaches can be categorized into two categories: 
(1) dynamic approaches which are based on running/executing the tests and (2) static approaches which are built on top of source code patterns or features.}

\vspace{0.2cm}
\noindent\textbf{\revised{Dynamic APCA approaches.}}   
Yang et al.~\cite{Yang2017BetterTC} leveraged fuzz strategies on existing test cases to automatically generate new test inputs.
Xin and Reiss~\cite{Xin2017IdentifyingTP} utilized the syntactic differences between the buggy code and its patched code to generate new test inputs. 
Xiong et al.~\cite{xiong2018identifying} focus on the behavior similarity of the failing tests on buggy and patched programs to assess the correctness of generated patches.
\revised{Ye et al.~\cite{Ye2021AutomatedPA} utilized a technique called ``Random testing with Ground Truth (RGT)''~\cite{shamshiri2015automatically} to
generate extra tests based on the human-written patch.}

\revised{
While dynamic approaches have shown promise, they tend to be highly time-consuming~\cite{Xin2017IdentifyingTP,Ye2021AutomatedPA}, whereas static approaches offer greater efficiency. Our approach falls within the category of static approaches. In this study, we aim to advance static APCA approaches.
We consider both static and dynamic approaches to be valuable research directions that contribute to progress in the field.}

\vspace{0.2cm}
\noindent\textbf{\revised{Static APCA approaches.}}  
Ye et al.~\cite{Ye2022AutomatedCO} proposed ODS to detect overfitting patches. 
They first statically extracted 4,199 code features at the AST level from the buggy code and generated patches by the APR tools.
Then they fed the extracted features to three machine learning algorithms (Logistic Regression,
k-Nearest Neighbors, and Random Forest) and ensemble the three models to assess the correctness.
Tian et al.~\cite{tian2020evaluating} leveraged representation learning techniques (e.g., BERT~\cite{bert}) to build embeddings for overfitting and correct patches generated by APR tools. They then fed the embeddings to machine learning classifiers (e.g. Logistic Regression) to obtain prediction results.
\revised{
Phung et al.~\cite{phung2022identifying} propose MIPI, an approach that leverages the similarity between the patched method's name (often indicative of the developer's intention) and the semantic meaning of the method's body (reflecting the implemented behavior) to detect and eliminate overfitting patches generated by APR tools.
}
\revised{
Ghanbari and Marcus~\cite{Ghanbari2022PatchCA} proposed Shibboleth, a method that measures the impact of patches on both production code and test code to separate the patches that result in similar programs. Shibboleth assesses the correctness of patches via both ranking and classification.
}
\revised{
Recently, Tian et al.~\cite{quatrain} introduced Quatrain, a static-based approach that redefines patch correctness evaluation as a question-answering task. Quatrain employs the natural language processing (NLP) technique to understand the relationship between bug reports and patch descriptions.
}
Lin et al.~\cite{cache} proposed Cache that utilized both the context and structure information in patches. Cache achieved state-of-the-art performances in the APCA task by outperforming existing dynamic and static APCA tools.
manually label the patches generated by a new/unseen APR tool, which indeed alleviates the manual labeling process for the APCA task.

\revised{
While \toolname utilizes the same labeled patches in the training set as other learning-based approaches, it distinguishes itself in the way it uses these patches. Previous methods like Cache~\cite{cache, Ye2022AutomatedCO, quatrain} incorporate these patches during model training, whereas our approach employs them during inference. We adopt this strategy due to the challenge of updating the parameters of large language models (LLMs), thus resorting to in-context learning to guide LLMs. Despite this difference in approach, our method doesn't necessitate additional labeled data or a correctness oracle. Hence, our comparison with baselines remains fair.
}

\vspace{0.2cm}
\noindent\textbf{\revised{Other Studies.}}
Recently, Le-Cong et al. ~\cite{le2023invalidator} proposed Invalidator that utilized both dynamic features (i.e., program invariants) and static features (i.e., code embedding extracted from CodeBERT). However, it is time-consuming to generate the dynamic features. Invalidator took five hours to infer dynamic features and seven minutes (on average) to assess the correctness for a single patch. While \toolname only costs 2.4 seconds for each patch.
Wang et al.~\cite{Wang2020AutomatedPC} performed a large-scale empirical study on the effectiveness of existing APCA approaches. 
\revised{Yang et al.~\cite{yang2023large} constructed a new APCA dataset and conducted an empirical study on the effectiveness of existing APCA approaches on the new dataset.}
\revised{
Motwani et al.~\cite{motwani2020quality} investigate the phenomenon in real-world Java programs, assessing the effectiveness of four program repair tools (GenProg, Par, SimFix, and TrpAutoRepair) on defects introduced by the projects' developers during their regular development process. They discovered that the generated patches frequently overfit to the provided test suite, with only 13.8\% to 46.1\% of the patches passing an independent set of tests.}
\revised{
Additionally, Xia et al.~\cite{xia2022less} introduced AlphaRepair, an infilling-style APR approach for generating bug fixes without the need for fine-tuning pre-trained models. While AlphaRepair shares similarities with our approach in not requiring fine-tuning, our focuses differ: Xia et al. concentrate on the program repair task, whereas we center our attention on patch correctness assessment. Furthermore, while we utilize large-size LLMs up to 7B parameters, Xia et al. only employ the small-size LLM CodeBERT with 0.13B parameters.
}

\section{Conclusion and Future Work}
\label{sec:conclusion}

\revised{In this study, we introduce \toolname, the first Large Language Models (LLMs) based automatic patch correctness assessment technique. 
\toolname utilizes an advanced LLM for code (Starcoder-7B) to assess the correctness of unlabeled patches of a
new or unseen APR tool, without the need for fine-tuning.
Moreover, \toolname incorporates a contrastive learning-based retrieval module to select similar patches as examples for each test patch, which helps the LLM better understand the correctness of the test patch.
Additionally, \toolname integrates diverse guiding information to enhance its decision-making process. Specifically, LLM4PatchCorrect incorporates bug descriptions, execution traces, failing test cases, and test coverage data.
Our experimental results showed that \toolname can achieve an accuracy of 84.4\% and an F1-score of 86.5\% on average although no labeled patch of the new or unseen APR tool is available.
In addition, \toolname significantly improves Accuracy, F1, and AUC scores, increasing them from 10.2\% to 32.4\%, 6.1\% to 24.1\%, and 10.1\% to 33.2\%, on average, respectively, compared to state-of-the-art approaches.}
For future work, we would like to explore the effectiveness of \toolname on other tasks such as just-in-time defect prediction.

\ifCLASSOPTIONcaptionsoff
  \newpage
\fi



%


\balance
\bibliographystyle{IEEEtran}
\bibliography{references}

%







\end{document}